\documentclass[runningheads]{llncs} 
\usepackage{iparg} 
\usepackage[hidelinks]{hyperref}
\hypersetup{colorlinks=true,linkcolor=blue,citecolor=blue,urlcolor=blue}
\usepackage{orcidlink}
\usepackage{version} 

\renewcommand{\doi}[1]{{doi:\href{https://doi.org/#1}{\nolinkurl{#1}}}}
\renewcommand{\url}[1]{{\href{#1}{\nolinkurl{#1}}}}

\title
 {Complementing an Imperative Process Algebra with a Rely/Guarantee Logic}
\author{C.A. Middelburg\,\orcidlink{0000-0002-8725-0197}}
\institute
 {Informatics Institute, Faculty of Science, University of Amsterdam \\
  Science Park~900, 1098~XH Amsterdam, the Netherlands \\
  \href{mailto:C.A.Middelburg@uva.nl}{C.A.Middelburg@uva.nl}}
  
\titlerunning
 {Complementing an Imperative Process Algebra with a Rely/Guarantee Logic}
\authorrunning
 {C.A. Middelburg}

\begin{document}
\maketitle

\begin{abstract}
This paper concerns the relation between imperative process algebra and 
rely/guarantee logic. 
An imperative process algebra is complemented by a rely/guarantee logic 
that can be used to reason about how data change in the course of a 
process. 
The imperative process algebra \linebreak[2] used is the extension of 
ACP (Algebra of Communicating Processes) that is used earlier in a paper 
about the relation between imperative process algebra and Hoare logic.
A complementing rely/guarantee logic that concerns judgments of partial 
correctness is treated in detail. 
The adaptation of this logic to weak and strong total correctness is 
also addressed.
A simple example is given that suggests that a rely/guarantee logic is 
more suitable as a complementing logic than a Hoare logic if interfering
parallel processes are involved.

\keywords{imperative process algebra \and rely/guarantee logic \and 
          asserted process \and partial correctness \and
          weak total correctness \and deadlock freedom}

\begin{classcode}
D.1.3, D.2.4, F.1.2, F.3.1 
\end{classcode}
\end{abstract}

\section{Introduction}
\label{sect-intro}

In~\cite{BM19b}, a process algebra is introduced that is an extension of 
ACP (Algebra of Communicating Processes) with features to communicate 
data between processes, to change data involved in a process in the 
course of the process, and to proceed at certain stages of a process in 
a way that depends on the changing data. 
The term imperative process algebra was coined in~\cite{NP97a} for 
process algebra with such features.
The process algebra in question provides a setting for equational 
reasoning about processes such as those carried out by many\linebreak[2] 
contemporary computer-based systems.
It is quite common that reasoning about such a process involves 
reasoning about how data involved in the process change in the course of 
the process.
In~\cite{BM19b}, the question whether and how a Hoare logic can be used 
for the latter kind of reasoning is addressed.

To keep this kind of reasoning compositional, a Hoare logic is 
introduced in~\cite{BM19b} that does not cover interfering parallel 
processes.
To deal with interference, reasoning with the axioms and rules of the 
Hoare logic has to be combined with reasoning with the equational axioms 
of the process algebra. 
In~\cite{Jon83a}, an approach to deal with the interference between 
parallel programs with a shared state in a compositional development 
method for concurrent programs is proposed.
The essence of this approach, the so-called rely/guarantee approach, has 
proved influential when it comes to compositional development of 
parallel programs with a shared state. 
There is extensive literature on Hoare-style logics that follows the 
rely/guarantee approach 
(see e.g.~\cite{Sti88a,Sto91a,XRH97a,CJ07a,WDP10a,Sta15a,SZLY21a}).
The question arises whether and how such a logic can be used in the 
setting of an imperative process algebra to reason about how data 
involved in a process change in the course of the process.

In the current paper, this question is addressed.
We develop a partial-correct\-ness Hoare-style logic for the processes 
considered in the imperative process algebra from~\cite{BM19b} that 
follows the rely/guarantee approach to deal with interference.
As an example, we describe the behaviour of a very simple system using 
the proces algebra and reason about how this system changes data with 
the axioms and rules of the logic developed.
The example suggests that a Hoare-style logic that follows the 
rely/guarantee approach is more suitable as a complementing logic than 
a Hoare logic if interfering parallel processes are involved.
We also address the adaptation of the logic to weak and strong total
correctness.

A Hoare-style logic that follows the rely/guarantee approach is 
increasingly referred to as a rely/guarantee logic.
In this paper, we follow this trend.
Like the judgments that can be derived in the Hoare logic developed 
in~\cite{BM19b}, the judgments that can be derived in the rely/guarantee
logic developed in the current paper are called asserted processes.
In the next paragraph, we will briefly outline the main differences 
between the asserted processes of the Hoare logic developed 
in~\cite{BM19b} and the asserted processes of the rely/guarantee logic 
developed in the current paper.
If the interference between parallel processes can be dealt with by 
means of the asserted processes of the latter logic, then the number of 
applications of axioms and rules required to derive an asserted process 
can potentially be reduced exponentially when using the latter logic.

In line with other Hoare logics, an asserted process of the Hoare logic 
developed in~\cite{BM19b} consists of a term that denotes a process, a 
pre-condition characterizing the set of data states to which the data 
state at the start of the process is assumed to belong, and a 
post-condition characterizing the set of data states to which the data 
state at successful termination of the process must belong.
In the rely/guarantee logic developed in the current paper, the asserted 
processes are augmented with a rely-condition characterizing the 
relation on data states that is assumed to hold between the data state 
before and the data state after each step taken by processes taking 
place in parallel with the process concerned and a guarantee-condition 
characterizing the relation on data states that must hold between the 
data state before and the data state after each step taken by the 
process concerned.

The interference between parallel processes cannot always be dealt with 
by means of rely- and guarantee-conditions.
Examples of their expressive weakness can be found, for instance, 
in~\cite{JP11a,JY19a,Yat24a}.
An at least partial technical explanation of the weakness is Lemma~6 
from~\cite{CJ07a}.
The more expressive alternatives known offer a combination of 
rely/guarantee ideas with a temporal logic 
(see e.g.~\cite{GCPV09a,STEPR14a}).
They trade tractability for expressiveness and sometimes do not even 
enforce compositionality.
Since compositionality and tractability are considered more important 
than expressiveness, it was decided not to consider a more expressive 
alternative.
Moreover, it is considered an added bonus that the rely/guarantee logic 
developed in the current paper is a natural generalization of the Hoare
logic developed in~\cite{BM19b}.

This paper is organized as follows.
First, the imperative process algebra used in this paper is presented 
(Sections~\ref{sect-ACPei}--\ref{sect-semantics}).
Then, a partial-correctness rely/guar\-antee logic for this process 
algebra is presented 
(Sections~\ref{sect-rely-guar}--\ref{sect-soundness-RG}).
After that, a simple example of the use of this logic is given 
(Section~\ref{sect-example}) 
and its adaptation to weak and strong correctness is addressed 
(Sections~\ref{sect-convergence}--\ref{sect-deadlock-freedom}).
Finally, some concluding remarks are made 
(Section~\ref{sect-conclusions}).

Sections~\ref{sect-ACPei} and~\ref{sect-deACPei} of this paper are 
abridged versions of Sections~2 and~3 of~\cite{BM19b}.
Portions of Sections~2 and~3 of that paper have been copied verbatim or 
slightly modified.

\section{\ACP\ with Empty Process and Iteration}
\label{sect-ACPei}

The imperative process algebra \deACPei\ is an extension of \ACPei, 
the version of \ACP\ that was first presented 
in~\cite[Section~4.4]{BW90}, extended with the binary iteration operator 
from~\cite{BBP94a}.%
\footnote
{In~\cite{BBP94a}, this iteration operator is called the binary Kleene
 star operator.}
In this section, a short survey of \ACPei\ is given.
In the next section, \deACPei\ is introduced as an extension of 
\ACPei.
After that, the soundness result for the axioms of \deACPei\ is 
presented.

In \ACPei, it is assumed that a fixed but arbitrary finite set $\Act$ 
of \emph{basic actions}, with $\dead,\ep \not\in \Act$, and a fixed 
but arbitrary commutative and associative \emph{communication} function 
$\funct{\commf}
 {(\Act \Sunion \Set{\dead}) \Sx (\Act \Sunion \Set{\dead})}
 {(\Act \Sunion \Set{\dead})}$, 
such that $\commf(\dead,a) = \dead$ 
for all $a \in \Act \Sunion \Set{\dead}$, have been given.
Basic actions are considered atomic processes.
The function $\commf$ is regarded to give the result of simultaneously
performing any two basic actions for which this is possible, and to be 
$\dead$ otherwise.
Henceforth, we write $\Actd$ for $\Act \Sunion \Set{\dead}$.

The algebraic theory \ACPei\ has one sort: the sort $\Proc$ of 
\emph{processes}.
This sort is made explicit to anticipate the need for many-sortedness 
later on. 
The algebraic theory \ACPei\ has the following constants and operators 
to build terms of sort~$\Proc$:
\begin{itemize}
\item
a \emph{basic action} constant $\const{a}{\Proc}$ for each 
$a \in \Act$;
\item
an \emph{inaction} constant $\const{\dead}{\Proc}$;
\item
an \emph{empty process} constant $\const{\ep}{\Proc}$;
\item
a binary \emph{alternative composition} or \emph{choice} operator 
$\funct{\altc}{\Proc \Sx \Proc}{\Proc}$;
\item
a binary \emph{sequential composition} operator 
$\funct{\seqc}{\Proc \Sx \Proc}{\Proc}$;
\item
a binary \emph{iteration} operator 
$\funct{\iter}{\Proc \Sx \Proc}{\Proc}$;
\item
a binary \emph{parallel composition} or \emph{merge} operator 
$\funct{\parc}{\Proc \Sx \Proc}{\Proc}$;
\item
a binary \emph{left merge} operator 
$\funct{\leftm}{\Proc \Sx \Proc}{\Proc}$;
\item
a binary \emph{communication merge} operator 
$\funct{\commm}{\Proc \Sx \Proc}{\Proc}$;
\item
a unary \emph{encapsulation} operator 
$\funct{\encap{H}}{\Proc}{\Proc}$ for each $H \subseteq \Act$.
\end{itemize}
It is assumed that there is a countably infinite set $\cX$ of variables 
of sort $\Proc$, which contains $x$, $y$ and $z$.
Terms are built as usual.
Infix notation is used for the binary operators.
The following precedence conventions are used to reduce the need for
parentheses: the operator ${} \seqc {}$ binds stronger than all other 
binary operators and the operator ${} \altc {}$ binds weaker than all 
other binary operators.

Let $t$ and $t'$ be closed \ACPei\ terms, and let $p$ and $p'$ be the 
processes denoted by $t$ and $t'$, respectively. 
Then the above constants and operators can be explained as follows:
\begin{itemize}
\item
$a$ denotes the process that first performs the action $a$ and then 
terminates successfully;
\item
$\ep$ denotes the process that terminates successfully without 
performing any action;
\item
$\dead$ denotes the process that cannot do anything, it cannot even 
terminate successfully;
\item
$t \altc t'$ denotes the process that behaves as either $p$ or $p'$;
\item
$t \seqc t'$\, denotes the process that behaves as $p$ and $p'$ in 
sequence;
\item
$t \iter t'$ denotes the process that first behaves as $p$ zero or more 
times and then behaves as $p'$;
\item
$t \parc t'$ denotes the process that behaves as $p$ and $p'$ in 
parallel;
\item
$t \leftm t'$ denotes the same process as $t \parc t'$, except that it 
starts with performing an action of $p$;
\item
$t \commm t'$ denotes the same process as $t \parc t'$, except that it 
starts with performing an action of $p$ and an action of $p'$ 
synchronously;
\item
$\encap{H}(t)$ denotes the process that behaves as $p$, except that 
actions from $H$ are blocked from being performed.
\end{itemize}

The axioms of \ACPei\ are presented in Table~\ref{axioms-ACPei}.
\begin{table}[!t]
\caption{Axioms of \ACPei}
\label{axioms-ACPei}
\begin{eqntbl}
\begin{axcol}
x \altc y = y \altc x                                & & \axiom{A1} \\
(x \altc y) \altc z = x \altc (y \altc z)            & & \axiom{A2} \\
x \altc x = x                                        & & \axiom{A3} \\
(x \altc y) \seqc z = x \seqc z \altc y \seqc z      & & \axiom{A4} \\
(x \seqc y) \seqc z = x \seqc (y \seqc z)            & & \axiom{A5} \\
x \altc \dead = x                                    & & \axiom{A6} \\
\dead \seqc x = \dead                                & & \axiom{A7} \\
x \seqc \ep = x                                      & & \axiom{A8} \\
\ep \seqc x = x                                      & & \axiom{A9} 
\eqnsep
x \parc y = x \leftm y \altc y \leftm x \altc x \commm y \altc
\encap{\Act}(x) \seqc \encap{\Act}(y)                & & \axiom{CM1E} \\
\ep \leftm x = \dead                                 & & \axiom{CM2E} \\
\alpha \seqc x \leftm y = \alpha \seqc (x \parc y)   & & \axiom{CM3}  \\
(x \altc y) \leftm z = x \leftm z \altc y \leftm z   & & \axiom{CM4}  \\
\ep \commm x = \dead                                 & & \axiom{CM5E} \\
x \commm \ep = \dead                                 & & \axiom{CM6E} \\
a \seqc x \commm b \seqc y = \commf(a,b) \seqc (x \parc y) 
                                                     & & \axiom{CM7}  \\
(x \altc y) \commm z = x \commm z \altc y \commm z   & & \axiom{CM8}  \\
x \commm (y \altc z) = x \commm y \altc x \commm z   & & \axiom{CM9}  
\eqnsep
\encap{H}(\ep) = \ep                                   & & \axiom{D0} \\
\encap{H}(a) = a                       & \mif a \notin H & \axiom{D1} \\ 
\encap{H}(a) = \dead                   & \mif a \in H    & \axiom{D2} \\
\encap{H}(x \altc y) = \encap{H}(x) \altc \encap{H}(y) & & \axiom{D3} \\
\encap{H}(x \seqc y) = \encap{H}(x) \seqc \encap{H}(y) & & \axiom{D4} 
\eqnsep
x \iter y = x \seqc (x \iter y) \altc y              & & \axiom{BKS1} \\
(x \altc \ep) \iter y = x \iter y                    & & \axiom{BKS5} \\
\encap{\Act}(x) = \dead \,\Land\, z = x \seqc z \altc y \;\Limpl\; 
                                       z = x \iter y & & \axiom{RSP*E}
\end{axcol}
\end{eqntbl}
\end{table}
In this table, $a$, $b$, and $\alpha$ stand for arbitrary members of 
$\Actd$,\, and $H$ stands for an arbitrary subset of $\Act$.
This means that CM3, CM7, and D0--D4 are in fact axiom schemas.
In this paper, axiom schemas will usually be referred to as axioms.
Axioms A1--A9, CM1E, CM2E, CM3, CM4, CM5E, CM6E, CM7--CM9, and D0--D4 
are the axioms of \ACPe\ (cf.~\cite{BW90}).
Axiom BKS1 is one of the axioms of \ACPi\ (cf.~\cite{BBP94a}) and axiom 
RSP*E is an adaptation of the axiom RSP* proposed in~\cite{BFP01a} to 
the presence of $\ep$ in \ACPei.
Axiom BKS5 is new.

The iteration operator originates from~\cite{BBP94a}, where it is called 
the binary Kleene star operator.
In the axiom system of \ACPi\ given in~\cite{BBP94a}, the axioms 
concerning the iteration operator are BKS1 and the equations BKS2--BKS4 
given in Table~\ref{deriv-eqns-ACPei}. 
\begin{table}[!t]
\caption{Derivable equations for iteration}
\label{deriv-eqns-ACPei}
\begin{eqntbl}
\begin{axcol}
x \iter (y \seqc z) = (x \iter y) \seqc z              & \axiom{BKS2} \\
x \iter (y \seqc ((x \altc y) \iter z) \altc z) = (x \altc y) \iter z
                                                       & \axiom{BKS3} \\
\encap{H}(x \iter y) = \encap{H}(x) \iter \encap{H}(y) & \axiom{BKS4} 
\end{axcol}
\end{eqntbl}
\end{table}
Equations BKS2--BKS4 are derivable from the axioms of \ACPei. 
Not all equations derivable from the axioms of \ACPei\ are also 
derivable from the axioms of \ACPei\ with RSP*E replaced by BKS2--BKS4.
Axiom BKS5 is not derivable from the other axioms of \ACPei.

Because conditional equational formulas must be dealt with in \ACPei, 
it is understood that conditional equational logic is used in deriving 
equations from the axioms of \ACPei.
A complete inference system for conditional equational logic can for
example be found in~\cite{BW90,Gog21a}.

\section{Imperative \ACPei}
\label{sect-deACPei}

In this section, \deACPei, imperative \ACPei, is introduced as an 
extension of \ACPei.
\deACPei\ extends \ACPei\ with features to communicate data between 
processes, to change data involved in a process in the course of the 
process, and to proceed at certain stages of a process in a way that 
depends on the changing data. 

In \deACPei, it is assumed that the following has been given with 
respect to data:
\begin{itemize}
\item
a many-sorted signature $\sign_\gD$ that includes:
\begin{itemize}
\item
a sort $\Data$ of \emph{data} and
a sort $\Bool$ of \emph{bits};
\item
constants of sort $\Data$ and/or operators with result sort $\Data$;
\item
constants $\zero$ and $\one$ of sort $\Bool$ and
operators with result sort $\Bool$;
\end{itemize}
\item
a minimal algebra $\gD$ of the signature $\sign_\gD$ in which 
the carrier of sort $\Bool$ has cardinality $2$ and 
the equation $\zero = \one$ does not hold.
\end{itemize}
It is moreover assumed that a finite or countably infinite set 
$\FlexVar$ of \emph{flexible variables} has been given.
A flexible variable is a variable whose value may change in the course 
of a process.%
\footnote
{The term flexible variable is used for this kind of variables in 
 e.g.~\cite{Sch97a,Lam94a}.} 
We write $\DataVal$ for the set of all closed terms over the signature
$\sign_\gD$ that are of sort $\Data$.
Moreover, we write $S^\gD$, where $S$ is a sort from $\sign_\gD$, for
$\gD$'s carrier of sort $S$.

A \emph{flexible variable valuation} is a function from $\FlexVar$ to 
$\DataVal$. 
Flexible variable valuations are intended to provide the values assigned to 
flexible variables when an \deACPei\ term of sort $\Data$ is evaluated.
To fit better in an algebraic setting, they provide closed terms from 
$\DataVal$ that denote those values instead.
Because $\gD$ is a minimal algebra, each member of $\Data^\gD$ can be 
represented by a member of~$\DataVal$. 

Below, the sorts, constants and operators of \deACPei\ are introduced.
The operators of \deACPei\ include a variable-binding operator.
\sloppy
The formation rules for \mbox{\deACPei}\ terms are the usual ones for 
the many-sorted case (see e.g.~\cite{ST99a,Wir90a}) and in addition the 
following rule:
\begin{itemize}
\item
if $O$ is a variable-binding operator 
$\funct{O}{S_1 \Sx \ldots \Sx S_n}{S}$ that binds a variable of sort~$S'$,
$t_1,\ldots,t_n$~are terms of sorts $S_1,\ldots,S_n$, respectively, and 
$X$ is a variable of sort $S'$, then $O X (t_1,\ldots,t_n)$ is a term of 
sort $S$.
\end{itemize}
An extensive formal treatment of the phenomenon of variable-binding 
operators can be found in~\cite{PS95a}.

\deACPei\ has the following sorts: 
the sorts included in $\sign_\gD$,
the sort $\Cond$ of \emph{conditions}, and
the sort $\Proc$ of \emph{processes}.

For each sort $S$ included in $\sign_\gD$ other than $\Data$, 
\deACPei\ has only the constants and operators included in $\sign_\gD$ 
to build terms of sort $S$.

\deACPei\ has, in addition to the constants and operators included in 
$\sign_\gD$ to build terms of sorts $\Data$, the following constants to 
build terms of sort $\Data$:
\begin{itemize}
\item
for each $v \in \FlexVar$, the \emph{flexible variable} constant 
$\const{v}{\Data}$.
\end{itemize}
We write $\DataTerm$ for the set of all closed \deACPei\ terms of sort 
$\Data$.

\deACPei\ has the following constants and operators to build terms of 
sort~$\Cond$:
\begin{itemize}
\item
a binary \emph{equality} operator
$\funct{=}{\Bool \Sx \Bool}{\Cond}$;
\item
a binary \emph{equality} operator
$\funct{=}{\Data \Sx \Data}{\Cond}$;%
\footnote
{The overloading of $=$ can be trivially resolved if $\sign_\gD$ is
 without overloaded symbols.}
\item
a \emph{falsity} constant $\const{\False}{\Cond}$;
\item
a unary \emph{negation} operator $\funct{\Lnot}{\Cond}{\Cond}$;
\item
a binary \emph{disjunction} operator 
$\funct{\Lor}{\Cond \Sx \Cond}{\Cond}$;
\item
a unary variable-binding \emph{existential quantification} operator 
$\funct{\exists}{\Cond}{\Cond}$ that binds a variable of sort $\Data$. 
\end{itemize}
We write $\CondTerm$ for the set of all closed \deACPei\ terms of sort 
$\Cond$.

\deACPei\ has, in addition to the constants and operators of \ACPei, 
the following operators to build terms of sort $\Proc$ ($n \in \Nat$):
\begin{itemize}
\item
an $n$-ary \emph{data parameterized action} operator
$\funct{a}{\Data ^n}{\Proc}$ for each $a \in \Act$;
\item
a unary \emph{assignment action} operator
$\funct{\assop{v}\,}{\Data}{\Proc}$ for each $v \in \FlexVar$;
\item
a binary \emph{guarded command} operator 
$\funct{\gc\,}{\Cond \Sx \Proc}{\Proc}$;
\item
a unary \emph{evaluation} operator 
$\funct{\eval{\rho}}{\Proc}{\Proc}$ for each $\rho \in \FVarVal$.
\end{itemize}
We write $\ProcTerm$ for the set of all closed \deACPei\ terms of sort 
$\Proc$.

It is assumed that there are countably infinite sets of variables of 
sort $\Data$ and $\Cond$ and that the sets of variables of sort $\Data$, 
$\Cond$, and $\Proc$ are mutually disjoint and disjoint from $\FlexVar$.

The same notational conventions are used as before.
Infix notation is also used for the additional binary operators.
Moreover, the notation $\ass{v}{e}$, where $v \in \FlexVar$ and $e$ is a 
\deACPei\ term of sort $\Data$, is used for the term $\assop{v}(e)$.

We also use the common logical abbreviations.
Let $\phi$ and $\psi$ be \deACPei\ terms of sort $\Cond$ and
let $X$ be a variable of sort $\Data$.
Then $\True$ stands for $\Lnot \False$,
$\phi \Land \psi$ stands for $\Lnot (\Lnot \phi \Lor \Lnot \psi)$,
$\phi \Limpl \psi$ stands for $\Lnot \phi \Lor \psi$,
$\phi \Liff \psi$ stands for 
$(\phi \Limpl \psi) \Land (\psi \Limpl \phi)$, and
$\Lforall{X}{\phi}$ stands for $\Lnot \Lexists{X}{\Lnot \phi}$.

Operators with result sort $\Bool$ serve as predicates.
Therefore, we usually write $p(e_1,\ldots,e_n)$, where 
$\funct{p}{\Data^n}{\Bool}$ and $e_1,\ldots,e_n$ are \deACPei\ terms of 
sort $\Data$, instead of  $p(e_1,\ldots,e_n) = 1$ where a term of sort 
$\Cond$ is expected.

Each term from $\CondTerm$ can be taken as a formula of a first-order 
language with equality of $\gD$ by taking the flexible variable 
constants as additional variables of sort $\Data$.
The flexible variable constants are implicitly taken as additional 
variables of sort $\Data$ wherever the context asks for a formula.
In this way, each term from $\CondTerm$ can be interpreted in $\gD$ as a
formula.
The axioms of \deACPei\ (given below) include an equation $\phi = \psi$ 
for each two terms $\phi$ and $\psi$ from $\CondTerm$ for which the 
formula $\phi \Liff \psi$ holds in $\gD$.

Let 
$a$ be a basic action from $\Act$, 
$e_1$, \ldots, $e_n$, and $e$ be terms from $\DataTerm$,  
$\phi$ be a term from $\CondTerm$, and 
$t$ be a term from $\ProcTerm$.
Then the additional operators to build terms of sort $\Proc$ can be 
explained as follows:
\begin{itemize}
\item
$a(e_1,\ldots,e_n)$ denotes the process that first performs the data 
parameterized action $a(e_1,\ldots,e_n)$ and then terminates 
successfully;
\item
$\ass{v}{e}$ denotes the process that first performs the assignment 
action $\ass{v}{e}$, whose intended effect is the assignment of the 
result of evaluating $e$ to flexible variable $v$, and then terminates 
successfully; 
\item
$\phi \gc t$ denotes the process that behaves as $p$ if condition $\phi$ 
holds for the values assigned to the flexible variables occurring in 
$\phi$ and as $\dead$ otherwise;
\item
$\eval{\rho}(t)$ denotes the process that remains of the process 
denoted by $t$ after evaluation of all subterms of $t$ that belong to 
$\DataTerm$ or $\CondTerm$ using flexible variable valuation $\rho$ 
updated according to the assignment actions performed.
\end{itemize}

A flexible variable valuation $\rho$ can be extended homomorphically 
from $\FlexVar$ to \deACPei\ terms of sort $\Data$ and \deACPei\ 
terms of sort $\Cond$.
Below, these extensions are denoted by $\rho$ as well.
Moreover, we write $\rho\mapupd{e}{v}$ for the flexible variable 
valuation $\rho'$ defined by $\rho'(v') = \rho(v')$ if $v' \neq v$ and 
$\rho'(v) = e$.

The subsets $\AProcPAR$, $\AProcASS$, and $\AProcTerm$ of $\ProcTerm$ 
referred to below are defined as follows:
\begin{ldispl}
\begin{aeqns}
\AProcPAR  & = & {} \Union_{n \in \Natpos}
\Set{a(e_1,\dots,e_n) \where
     a \in \Act \Land
     e_1,\dots,e_n \in \DataTerm}\;, \\
\AProcASS & = & 
\Set{\ass{v}{e} \where
     v \in \FlexVar \Land e \in \DataTerm}\;, \\
\AProcTerm & = & \Act \Sunion \AProcPAR \Sunion \AProcASS\;. 
\end{aeqns}
\end{ldispl}%
The elements of $\AProcTerm$ are the terms from $\ProcTerm$ that denote 
the processes that are considered atomic.
We write $\AProcTermd$ for $\AProcTerm \Sunion \Set{\dead}$.

The axioms of \deACPei\ are the axioms presented in 
Tables~\ref{axioms-ACPei} and~\ref{axioms-deACPei},
\begin{table}[!t]
\caption{Additional axioms of \deACPei}
\label{axioms-deACPei}
\begin{eqntbl}
\begin{axcol}
e = e'         & \mif \Sat{\gD}{\fol{e = e'}}          & \axiom{IMP1} \\
\phi = \psi    & \mif \Sat{\gD}{\fol{\phi \Liff \psi}} & \axiom{IMP2} 
\eqnsep
\True \gc x = x                                      & & \axiom{GC1}  \\
\False \gc x = \dead                                 & & \axiom{GC2}  \\
\phi \gc \dead = \dead                               & & \axiom{GC3}  \\
\phi \gc (x \altc y) = \phi \gc x \altc \phi \gc y   & & \axiom{GC4}  \\
\phi \gc x \seqc y = (\phi \gc x) \seqc y            & & \axiom{GC5}  \\
\phi \gc (\psi \gc x) = (\phi \Land \psi) \gc x      & & \axiom{GC6}  \\
(\phi \Lor \psi) \gc x = \phi \gc x \altc \psi \gc x & & \axiom{GC7}  \\
(\phi \gc x) \leftm y = \phi \gc (x \leftm y)        & & \axiom{GC8}  \\
(\phi \gc x) \commm y = \phi \gc (x \commm y)        & & \axiom{GC9}  \\
x \commm (\phi \gc y) = \phi \gc (x \commm y)        & & \axiom{GC10} \\
\encap{H}(\phi \gc x) = \phi \gc \encap{H}(x)        & & \axiom{GC11} 
\eqnsep
\eval{\rho}(\ep) = \ep                               & & \axiom{V0}   \\
\eval{\rho}(\alpha \seqc x) = \alpha \seqc \eval{\rho}(x)
      & \mif \alpha \notin \AProcPAR \Sunion \AProcASS & \axiom{V1}   \\
\multicolumn{2}{@{}l@{}}{
\eval{\rho}(a(e_1,\ldots,e_n) \seqc x) = 
a(\rho(e_1),\ldots,\rho(e_n)) \seqc \eval{\rho}(x)}    & \axiom{V2}   \\
\eval{\rho}(\ass{v}{e} \seqc x) = 
\ass{v}{\rho(e)} \seqc \eval{\rho\mapupd{\rho(e)}{v}}(x) 
                                                     & & \axiom{V3}   \\ 
\eval{\rho}(x \altc y) = \eval{\rho}(x) \altc \eval{\rho}(y)
                                                     & & \axiom{V4}   \\
\eval{\rho}(\phi \gc y) = \rho(\phi) \gc \eval{\rho}(x)          
                                                     & & \axiom{V5}   
\eqnsep
a(e_1,\ldots,e_n) \seqc x \commm b(e'_1,\ldots,e'_n) \seqc y = 
 {} \\ \quad
(e_1 = e'_1 \Land \ldots \Land e_n = e'_n) \gc c(e_1,\ldots,e_n) \seqc
(x \parc y)\;                  & \mif \commf(a,b) = c & \axiom{CM7Da} \\
a(e_1,\ldots,e_n) \seqc x \commm b(e'_1,\ldots,e'_m) \seqc y = \dead
& \mif \commf(a,b) = \dead \;\mathrm{or}\; n \neq m\; & \axiom{CM7Db} \\
a(e_1,\ldots,e_n) \seqc x \commm \alpha \seqc y = \dead 
& \mif \alpha \notin \AProcPAR                        & \axiom{CM7Dc} \\
\alpha \seqc x \commm a(e_1,\ldots,e_n) \seqc y = \dead 
& \mif \alpha \notin \AProcPAR                        & \axiom{CM7Dd} \\
\ass{v}{e} \seqc x \commm \alpha \seqc y = \dead    & & \axiom{CM7De} \\
\alpha \seqc x \commm \ass{v}{e} \seqc y = \dead    & & \axiom{CM7Df}  
\eqnsep
\encap{H}(a(e_1,\ldots,e_n)) = a(e_1,\ldots,e_n)
                                    & \mif a \notin H & \axiom{D1Da}  \\ 
\encap{H}(a(e_1,\ldots,e_n)) = \dead            
                                    & \mif a \in H    & \axiom{D2D}   \\
\encap{H}(\ass{v}{e}) = \ass{v}{e}                  & & \axiom{D1Db}
\end{axcol}
\end{eqntbl}
\end{table}
where
$\alpha$ stands for an arbitrary term from $\AProcTermd$,\, 
$H$ stands for an arbitrary subset of $\Act$,\, \linebreak[2]
$e$, $e_1,e_2,\ldots$\ and $e'$, $e'_1,e'_2,\ldots$\ stand for 
arbitrary terms from $\DataTerm$,\, 
$\phi$ and $\psi$ stand for arbitrary terms from $\CondTerm$,
$v$ stands for an arbitrary flexible variable from $\FlexVar$, and
$\rho$ stands for an arbitrary flexible variable valuation from 
$\FVarVal$.
Moreover,\, $a$, $b$, and $c$ stand for arbitrary members of $\Actd$ in 
Table~\ref{axioms-ACPei} and for arbitrary members of $\Act$ in 
Table~\ref{axioms-deACPei}.

All closed \deACPei\ terms of sort $\Proc$ can be brought into a 
particular form, the so-called the head normal form.

The set $\HNF$ of \emph{head normal forms} of \deACPei\ is inductively 
defined by the following rules:
\begin{itemize}
\item 
$\dead \in \HNF$;
\item 
if $\phi \in \CondTerm$, then $\phi \gc \ep \in \HNF$;
\item 
if $\phi \in \CondTerm$, $\alpha \in \AProcTerm$, and $p \in \ProcTerm$, 
then $\phi \gc \alpha \seqc p \in \HNF$;
\item 
if $p,p' \in \HNF$, then $p \altc p' \in \HNF$.
\end{itemize}
\begin{lemma}
\label{lemma-HNF}
For all terms $p \in \ProcTerm$, there exists a term $q \in \HNF$ such 
that $p = q$ is derivable from the axioms of \deACPei.
\end{lemma}
\begin{proof}
This is straightforwardly proved by induction on the structure of $p$.
The cases where $p$ is of the form $\dead$, $\ep$ or $\alpha$
($\alpha \in \AProcTerm$) are trivial.
The case where $p$ is of the form $p_1 \altc p_2$ follows immediately
from the induction hypothesis. 
The case where $p$ is of the form $p_1 \parc p_2$ follows immediately
from the case that $p$ is of the form $p_1 \leftm p_2$ and the case that
$p$ is of the form $p_1 \commm p_2$.
Each of the other cases follow immediately from the induction hypothesis
and a claim that is easily proved by structural induction.
In the case where $p$ is of the form $p_1 \commm p_2$, each of the cases
to be considered in the inductive proof demands an additional proof by
structural induction.
\qed
\end{proof}

Let $t$ and $t'$ be \deACPei\ terms of sort $\Proc$.
Then \emph{$t$ is a summand of $t'$}, written $t \sqsubseteq t'$, iff
there exists a \deACPei\ term $t''$ of sort $\Proc$ such that
$t \altc t'' = t'$ is derivable from axioms A1 and A2 or $t = t'$ is 
derivable from axioms A1 and A2. 

A result relating the notion of a summand defined above to the semantics 
of \deACPei\ will be given in Section~\ref{sect-semantics}.

\section{Bisimulation Semantics of \deACPei}
\label{sect-semantics}

In this section, a structural operational semantics of \deACPei\ is 
presented and a notion of bisimulation equivalence for \deACPei\ based 
on this structural operational semantics is defined.

\pagebreak[2]
The structural operational semantics of \deACPei\ consists of 
\nopagebreak[2]
\begin{itemize}
\item 
a binary \emph{transition} relation 
\smash{$\step{\gact{\rho}{\alpha}}$} on $\ProcTerm$ for each 
$\rho \in \FVarVal$ and $\alpha \in \AProcTerm$;
\item 
a unary \emph{successful termination} relation $\sterm{\rho}$ on 
$\ProcTerm$ for each $\rho \in \FVarVal$.
\end{itemize}
We write \smash{$\astep{t}{\gact{\rho}{\alpha}}{t'}$} instead of 
\smash{$(t,t') \in {\step{\gact{\rho}{\alpha}}}$} and 
$\isterm{t}{\rho}$ instead of $t \in {\sterm{\rho}}$.

The relations from the structural operational semantics describe what 
the processes denoted by terms from $\ProcTerm$ are capable of doing as 
follows:
\begin{itemize}
\item
$\astep{t}{\gact{\rho}{\alpha}}{t'}$: 
if the value of the flexible variables are as defined by $\rho$, then 
the process denoted by $t$ has the potential to make a transition to the 
process denoted by $t'$ by performing action $\alpha$;
\item
$\isterm{t}{\rho}$: 
if the value of the flexible variables are as defined by $\rho$, then 
the process denoted by $t$ has the potential to terminate successfully.
\end{itemize}

The relations from the structural operational semantics of \deACPei\ 
are the smallest relations satisfying the rules given in 
Table~\ref{sos-deACPei}.%
\begin{table}[!p]
\caption{Transition rules for \deACPei}
\label{sos-deACPei}
\begin{ruletbl}
{} \\[-3ex]
\Rule
{}
{\astep{\alpha}{\gact{\rho}{\alpha}}{\ep}}
\\[-2ex]
\Rule
{\phantom{\isterm{\ep}{\rho}}}
{\isterm{\ep}{\rho}}
\\
\Rule
{\isterm{x}{\rho}}
{\isterm{x \altc y}{\rho}}
\quad
\Rule
{\isterm{y}{\rho}}
{\isterm{x \altc y}{\rho}}
\quad
\Rule
{\astep{x}{\gact{\rho}{\alpha}}{x'}}
{\astep{x \altc y}{\gact{\rho}{\alpha}}{x'}}
\quad
\Rule
{\astep{y}{\gact{\rho}{\alpha}}{y'}}
{\astep{x \altc y}{\gact{\rho}{\alpha}}{y'}}
\\
\Rule
{\isterm{x}{\rho},\; \isterm{y}{\rho}}
{\isterm{x \seqc y}{\rho}}
\quad
\Rule
{\isterm{x}{\rho},\; \astep{y}{\gact{\rho}{\alpha}}{y'}}
{\astep{x \seqc y}{\gact{\rho}{\alpha}}{y'}}
\quad
\Rule
{\astep{x}{\gact{\rho}{\alpha}}{x'}}
{\astep{x \seqc y}{\gact{\rho}{\alpha}}{x' \seqc y}}
\\
\Rule
{\isterm{y}{\rho}}
{\isterm{x \iter y}{\rho}}
\quad
\Rule
{\astep{y}{\gact{\rho}{\alpha}}{y'}}
{\astep{x \iter y}{\gact{\rho}{\alpha}}{y'}}
\quad
\Rule
{\astep{x}{\gact{\rho}{\alpha}}{x'}}
{\astep{x \iter y}{\gact{\rho}{\alpha}}{x' \seqc (x \iter y)}}
\\
\Rule
{\isterm{x}{\rho},\; \isterm{y}{\rho}}
{\isterm{x \parc y}{\rho}}
\quad
\Rule
{\astep{x}{\gact{\rho}{\alpha}}{x'}}
{\astep{x \parc y}{\gact{\rho}{\alpha}}{x' \parc y}}
\quad
\Rule
{\astep{y}{\gact{\rho}{\alpha}}{y'}}
{\astep{x \parc y}{\gact{\rho}{\alpha}}{x \parc y'}}
\\
\RuleC
{\astep{x}{\gact{\rho}{a}}{x'},\; \astep{y}{\gact{\rho}{b}}{y'}}
{\astep{x \parc y}{\gact{\rho}{c}}{x' \parc y'}}
{\commf(a,b) = c}
\\
\RuleC
{\astep{x}{\gact{\rho}{a(e_1,\ldots,e_n)}}{x'},\; 
 \astep{y}{\gact{\rho}{b(e'_1,\ldots,e'_n)}}{y'}}
{\astep{x \parc y}{\gact{\rho}{c(e_1,\ldots,e_n)}}{x' \parc y'}}
{\!\commf(a,b) = c,\,
 \Sat{\gD}{\rho(\fol{e_1 = e'_1 \Land \ldots \Land  e_n = e'_n})}}
\\
\Rule
{\astep{x}{\gact{\rho}{\alpha}}{x'}}
{\astep{x \leftm y}{\gact{\rho}{\alpha}}{x' \parc y}}
\\
\RuleC
{\astep{x}{\gact{\rho}{a}}{x'},\; \astep{y}{\gact{\rho}{b}}{y'}}
{\astep{x \commm y}{\gact{\rho}{c}}{x' \parc y'}}
{\commf(a,b) = c}
\\
\RuleC
{\astep{x}{\gact{\rho}{a(e_1,\ldots,e_n)}}{x'},\; 
 \astep{y}{\gact{\rho}{b(e'_1,\ldots,e'_n)}}{y'}}
{\astep{x \commm y}
  {\gact{\rho}{c(e_1,\ldots,e_n)}}{x' \parc y'}}
{\!\commf(a,b) = c,\,
 \Sat{\gD}{\rho(\fol{e_1 = e'_1 \Land \ldots \Land  e_n = e'_n})}}
\\
\Rule
{\isterm{x}{\rho}}
{\isterm{\encap{H}(x)}{\rho}}
\quad
\RuleC
{\astep{x}{\gact{\rho}{a}}{x'}}
{\astep{\encap{H}(x)}{\gact{\rho}{a}}{\encap{H}(x')}}
{a \notin H}
\quad
\RuleC
{\astep{x}{\gact{\rho}{a(e_1,\ldots,e_n)}}{x'}}
{\astep{\encap{H}(x)}{\gact{\rho}{a(e_1,\ldots,e_n)}}{\encap{H}(x')}}
{a \notin H}
\\
\Rule
{\astep{x}{\gact{\rho}{\ass{v}{e}}}{x'}}
{\astep{\encap{H}(x)}{\gact{\rho}{\ass{v}{e}}}{\encap{H}(x')}}
\\
\RuleC
{\isterm{x}{\rho}}
{\isterm{\phi \gc x}{\rho}}
{\Sat{\gD}{\rho(\fol{\phi})}}
\quad
\RuleC
{\astep{x}{\gact{\rho}{\alpha}}{x'}}
{\astep{\phi \gc x}{\gact{\rho}{\alpha}}{x'}}
{\Sat{\gD}{\rho(\fol{\phi})}}
\\
\Rule
{\isterm{x}{\rho}}
{\isterm{\eval{\rho}(x)}{\rho'}}
\quad
\Rule
{\astep{x}{\gact{\rho}{a}}{x'}}
{\astep{\eval{\rho}(x)}{\gact{\rho'}{a}}{\eval{\rho}(x')}}
\quad
\Rule
{\astep{x}{\gact{\rho}{a(e_1,\ldots,e_n)}}{x'}}
{\astep{\eval{\rho}(x)}
       {\gact{\rho'}{a(\rho(e_1),\ldots,\rho(e_n))}}
       {\eval{\rho}(x')}}
\\
\Rule
{\astep{x}{\gact{\rho}{\ass{v}{e}}}{x'}}
{\astep{\eval{\rho}(x)}{\gact{\rho'}{\ass{v}{\rho(e)}}}
       {\eval{\rho\mapupd{\rho(e)}{v}}(x')}}
\vspace*{1ex}
\end{ruletbl}
\end{table}
In this table, 
$\rho$ and $\rho'$ stand for arbitrary flexible variable valuations from 
$\FVarVal$,\,
$\alpha$ stands for an arbitrary action from~$\AProcTerm$,\,
$a,b$, and $c$ stand for arbitrary basic actions from $\Act$,\,
$e,e_1,e_2,\ldots$ and $e'_1,e'_2,\ldots$ stand for arbitrary terms 
from~$\DataTerm$,\,
$H$ stands for an arbitrary subset of $\Act$,\,
$\phi$ stands for an arbitrary term from $\CondTerm$,\, and
$v$ stands for an arbitrary flexible variable from $\FlexVar$.

The rules in Table~\ref{sos-deACPei} have the form 
\smash{\small$\SRuleC{p_1,\ldots,p_n}{\raisebox{.6ex}{$c$}}{s}$}, where 
$s$ is optional.
They are to be read as ``if $p_1$ and \ldots and $p_n$ then $c$, 
provided $s$''.
As usual, $p_1,\ldots,p_n$ are called the premises and $c$ is called the 
conclusion.
A side condition $s$, if present, serves to restrict the applicability 
of a rule.
If a rule has no premises, then nothing is displayed above the 
horizontal bar.

Because the rules in Table~\ref{sos-deACPei} constitute an inductive 
definition, \mbox{$\astep{t}{\gact{\rho}{\alpha}}{t'}$} or 
$\isterm{t}{\rho}$ holds iff it can be inferred from these rules.

Two processes are considered equal if they can simulate each other 
insofar as their observable potentials to make transitions and to 
terminate successfully are concerned, taking into account the assigments 
of values to flexible variables under which the potentials are available.
This can be dealt with by means of the notion of bisimulation 
equivalence introduced in~\cite{GW96a} adapted to the conditionality of 
transitions.

An equivalence relation on the set $\AProcTerm$ is needed.
Two actions $\alpha,\alpha' \in \AProcTerm$ are \emph{data equivalent}, 
written $\alpha \simeq \alpha'$, iff one of the following holds:
\begin{itemize}
\item
there exists an $a \in \Act$ such that $\alpha = a$ 
and $\alpha' = a$;
\item
for some $n \in \Natpos$,
there exist an $a \in \Act$ and 
$e_1,\dots,e_n,e'_1,\dots,e'_n \in \DataTerm$
such that 
$\Sat{\gD}{\fol{e_1 = e'_1}}$, \ldots, $\Sat{\gD}{\fol{e_n = e'_n}}$,
$\alpha = a(e_1,\dots,e_n)$, and $\alpha' = a(e'_1,\dots,e'_n)$;
\item
there exist a $v \in \FlexVar$ and $e,e' \in \DataTerm$ such that 
$\Sat{\gD}{\fol{e = e'}}$, $\alpha = \ass{v}{e}$, and 
$\alpha' = \ass{v}{e'}$.
\end{itemize}
We write $[\alpha]$, where $\alpha \in \AProcTerm$, for the equivalence 
class of $\alpha$ with respect to $\simeq$.

A \emph{bisimulation} is a binary relation $R$ on $\ProcTerm$ such that, 
for all terms $t_1,t_2 \in \ProcTerm$ with $(t_1,t_2) \in R$, the 
following \emph{transfer conditions} hold:
\begin{itemize}
\item
if $\astep{t_1}{\gact{\rho}{\alpha}}{t_1'}$, then there exist an 
$\alpha' \in [\alpha]$ and a $t_2' \in \ProcTerm$ such that 
$\astep{t_2}{\gact{\rho}{\alpha'}}{t_2'}$ and 
$(t_1',t_2') \in R$;
\item
if $\astep{t_2}{\gact{\rho}{\alpha}}{t_2'}$, then there exist an 
$\alpha' \in [\alpha]$ and a $t_1' \in \ProcTerm$ such that 
$\astep{t_1}{\gact{\rho}{\alpha'}}{t_1'}$ and 
$(t_1',t_2') \in R$;
\item
if $\isterm{t_1}{\rho}$, then  $\isterm{t_2}{\rho}$;
\item
if $\isterm{t_2}{\rho}$, then  $\isterm{t_1}{\rho}$.
\end{itemize}

Two terms $t_1,t_2 \in \ProcTerm$ are \emph{bisimulation equivalent}, 
written $t_1 \bisim t_2$, if there exists a bisimulation $R$ such that 
$(t_1,t_2) \in R$.
Let $R$ be a bisimulation such that $(t_1,t_2) \in R$.
Then we say that $R$ is a bisimulation \emph{witnessing}
$t_1 \bisim t_2$.

Below, some results about bisimulation equivalence are given.
The routine proofs of these results are outlined in 
Appendix~\ref{appendix-proofs}.

Bisimulation equivalence is an equivalence relation.%
\begin{proposition}[Equivalence]
\label{proposition-equiv-deACPei}
The relation $\bisim$ is an equivalence relation. 
\end{proposition}
Moreover, bisimulation equivalence is a congruence with respect to the 
operators of \deACPei\ of which the result sort and at least one 
argument sort is $\Proc$.
\begin{proposition}[Congruence]
\label{proposition-congr-deACPei}
For all terms $t_1,t_1',t_2,t_2' \in \ProcTerm$ and all terms 
$\phi \in \CondTerm$, 
$t_1 \bisim t_2$ and $t_1' \bisim t_2'$ only if 
$t_1 \altc t_1' \bisim t_2 \altc t_2'$, 
$t_1 \seqc t_1' \bisim t_2 \seqc t_2'$, 
$t_1 \iter t_1' \bisim t_2 \iter t_2'$, 
$t_1 \parc t_1' \bisim t_2 \parc t_2'$, 
$t_1 \leftm t_1' \bisim t_2 \leftm t_2'$,
$t_1 \commm t_1' \bisim t_2 \commm t_2'$,
$\encap{H}(t_1) \bisim \encap{H}(t_2)$, 
$\phi \gc t_1 \bisim \phi \gc t_2$, and
$\eval{\sigma}(t_1) \bisim \eval{\sigma}(t_2)$.
\end{proposition}

The axiom system of \deACPei\ is sound with respect to bisimulation 
equivalence for equations between terms from $\ProcTerm$.
\begin{theorem}[Soundness]
\label{theorem-soundness-deACPei}
For all terms $t,t' \in \ProcTerm$, $t = t'$ is derivable from the 
axioms of \deACPei\ only if $t \bisim t'$.
\end{theorem}

We have been unable to provide a proof of the completeness of the axiom
system of \deACPei\ with respect to bisimulation equivalence for 
equations between terms from $\ProcTerm$.
In 1984, Milner posed in~\cite{Mil84a} the question whether the axiom 
system of the variant of \ACPei\ without the operators $\parc$, 
$\leftm$, $\commm$, and $\encap{H}$ in which the binary iteration 
operator is replaced by a unary iteration operator is complete.
It took several attempts by different scientists before, after 38 years, 
an affirmative answer was outlined by Grabmayer in~\cite{Grab22a}.
Because his proof is very complex, a monograph about the details of the 
completeness proof in question is currently being written.  
At best, the proof of the completeness of the axiom system of \deACPei\ 
will be a relatively simple adaptation of that proof.

The following result relates the summand relation $\sqsubseteq$ to the 
structural operational semantics of \deACPei.
\begin{lemma}
\label{lemma-summand}
For all terms $\alpha \in \AProcTerm$, $t' \in \ProcTerm$, and 
$t \in \HNF$, $\True \gc \alpha \seqc t' \sqsubseteq t$ iff
for all $\rho \in \FVarVal$, 
\smash{$\astep{t}{\gact{\rho}{\alpha}}{t'}$}.
\end{lemma}
\begin{proof}
Both the if part and the only-if part follow directly from the
definition of $\sqsubseteq$ and the structural operational semantics of
\deACPei.
\qed
\end{proof}

\section{A Rely/Garantee Logic of Asserted Processes}
\label{sect-rely-guar}

In this section, we present \RG, a rely/guarantee logic of asserted 
processes based on \deACPei, define what it means that an asserted 
process is true in the sense of partial correctness, and show that the 
axioms and rules of this logic are sound with respect to this meaning.

We write $\RGProcTerm$ for the set of all closed terms of sort $\Proc$ 
in which the evaluation operators $\eval{\sigma}$ and the auxiliary 
operators $\leftm$ and $\commm$ do not occur.
Moreover, we write $\stRGCondTerm$ for the set of all terms of sort 
$\Cond$ in which variables of sort $\Cond$ do not occur and 
$\RGDataTerm$ for the set of all terms of sort $\Data$.
Clearly, $\RGProcTerm \subset \ProcTerm$, 
$\CondTerm \subset \stRGCondTerm$, and $\DataTerm \subset \RGDataTerm$. 

In addition to the conditions from $\stRGCondTerm$, there is a need for 
conditions that allow relating the values of flexible variables 
immediately before and after a step of a process.
In the conditions in question, for each flexible variable $v$, two 
flexible variable constants can be used: the (current-value) flexible 
variable constant $v$ and an additional previous-value flexible variable 
constant $\prevvar{v}$.
Two flexible variable valuations will be used to evaluated these 
conditions.

Because of the need for the above-mentioned conditions, we extend the 
set of all terms of sort $\Cond$ by adding the following constants to 
the constants of sort~$\Data$: 
\begin{itemize}
\item
for each $v \in \FlexVar$, the \emph{previous-value flexible variable} 
constant $\const{\prevvar{v}}{\Data}$.
\end{itemize}
We write $\trRGCondTerm$ for the set of all terms from this set in which 
variables of sort $\Cond$ do not occur.%
\footnote{The subscript $\mathit{lb}$ abbreviates looking back.}
Clearly, $\stRGCondTerm \subset \trRGCondTerm$.
Moreover, we write \smash{$\prevvar{\FlexVar}$} for the set 
$\Set{\prevvar{v} \where v \in \FlexVar}$.

Let $\rho$ and $\rho'$ be flexible variable valuations.
Then the function 
$\funct{[\rho,\rho']}{\FlexVar \Sunion \prevvar{\FlexVar}}{\DataVal}$ is
defined as follows ($v \in \FlexVar$):
\begin{ldispl}
[\rho,\rho'](v)           = \rho(v)\;,  \quad
[\rho,\rho'](\prevvar{v}) = \rho'(v)\;.
\end{ldispl}%
A function $[\rho,\rho']$, where $\rho$ and $\rho'$ are flexible 
variable valuations, can be extended homomorphically 
from $\FlexVar$ to \deACPei\ terms of sort $\Data$ and \deACPei\ 
terms of sort $\Cond$.
Below, these extensions are denoted by $[\rho,\rho']$ as well.

An \emph{asserted process} is a judgment of the form 
$\assproc{R}{G}{\phi}{p}{\psi}$, where $p \in \RGProcTerm$,
$\phi,\psi \in \stRGCondTerm$, and $R,G \in \trRGCondTerm$.  
Here, $\phi$, $\psi$, $R$, and $G$ are called the \emph{pre-condition}, 
the \emph{post-condition}, the \emph{rely-condition}, and
the \emph{guarantee-condition}, respectively, of the asserted process.

Informally, an asserted process $\assproc{R}{G}{\phi}{p}{\psi}$ is true 
in the sense of partial \linebreak[2] correctness if the following is 
the case:
\begin{quote}
if $\phi$ holds at the start of $p$ and $R$ holds for each step taken by
processes taking place in parallel with $p$, then 
$G$ holds for each step taken by $p$ and, if $p$ eventually terminates 
successfully, $\psi$ holds upon successful termination of $p$.
\end{quote}
The conditions $\phi$ and $\psi$ concern the data values assigned to 
flexible variables at the start and at successful termination, 
respectively.
Because variables of sort $\Data$ may occur in $\phi$ and $\psi$, it is 
possible to refer in $\psi$ to the data values assigned to flexible 
variables at the start.

An asserted process $\assproc{\False}{\True}{\phi}{p}{\psi}$ concerns a
process considered in isolation.
The rely condition $\False$ indicates that processes taking place in 
parallel with $p$ can take no step and the guarantee condition $\True$ 
indicates that $p$ can take whatever step.

Below will be defined what it means that an asserted process is true in 
the sense of partial correctness.
First, some auxiliary notions will be defined.

The following auxiliary relation will be used:
\begin{itemize}
\item 
a ternary \emph{step} relation 
${\step{}} \subseteq
 (\RGProcTerm \Sx (\FVarVal)) \Sx (\AProcTerm \Sunion \Set{\sfe}) \Sx
 (\RGProcTerm \Sx (\FVarVal))$.
\end{itemize}
We write \smash{$(p,\rho) \step{l} (p',\rho')$} instead of 
$((p,\rho),l,(p',\rho')) \in {\step{}}$.

The step relation $\step{}$ is defined as follows: 
\begin{ldispl}
(p,\rho) \step{\alpha} (p',\rho') \;\;\text{iff}\;\;
\text{for all}\; \rho'' \in \FVarVal,\; 
\astep{\eval{\rho}(p)}{\gact{\rho''}{\alpha}}{\eval{\rho'}(p')}\;,
\\
(p,\rho) \estep (p,\rho')\;,
\end{ldispl}%
for all $p,p' \in \RGProcTerm$ and $\rho,\rho' \in \FVarVal$.

The step relation $\step{}$ can be explained as follows:
\begin{itemize}
\item
$(p,\rho) \step{\alpha} (p',\rho')$: 
if the value of the flexible variables are as defined by $\rho$, then 
$p$ can make a step by performing $\alpha$ and after this step $p$ has 
evolved to $p'$ and the values of the flexible variables are as defined 
by $\rho'$;
\item
$(p,\rho) \estep (p,\rho')$: 
if the value of the flexible variables are as defined by $\rho$, then 
the environment can make a step by performing some
$\alpha \in \AProcTerm$ and after this step $p$ has not evolved but the 
values of the flexible variables are as defined by $\rho'$.
\end{itemize}

Let $p \in \RGProcTerm$.
Then a \emph{computation of $p$} is a sequence
$((p_1,\rho_1), l_1, (p_2,\rho_2)),\linebreak[2] \ldots,
 ((p_n,\rho_n), l_n, (p_{n+1},\rho_{n+1}))$
over 
$(\RGProcTerm \Sx (\FVarVal)) \Sx (\AProcTerm \Sunion \Set{\sfe}) \Sx
 (\RGProcTerm \Sx (\FVarVal))$,
such that $p_1 = p$ and, 
for each $((p_i,\rho_i), l_i, (p_{i+1},\rho_{i+1}))$,\,
\smash{$(p_i,\rho_i) \step{l_i} (p_{i+1},\rho_{i+1})$}.

Let $p \in \RGProcTerm$, $\phi,\psi \in \stRGCondTerm$, 
and $R,G \in \trRGCondTerm$ and let
$\sigma = 
 ((p_1,\rho_1), l_1, (p_2,\rho_2)),\linebreak[2] \ldots,
 ((p_n,\rho_n), l_n, (p_{n+1},\rho_{n+1}))$
be a computation of $p$.
Then $\sigma$ \emph{satisfies the assumptions $\phi$ and $R$}, written
$\sigma \asat (\phi,R)$, iff
\begin{itemize}
\item
$\Sat{\gD}{\rho_1(\phi)}$;
\item
for each $((p_i,\rho_i), l_i, (p_{i+1},\rho_{i+1}))$ with $l_i = \sfe$,
$\Sat{\gD}{[\rho_{i+1},\rho_i](R)}$ 
\end{itemize}
and $\sigma$ \emph{satisfies the commitments $\psi$ and $G$}, written
$\sigma \csat (\psi,G)$, iff
\begin{itemize}
\item
$\Sat{\gD}{\rho_{n+1}(\psi)}$ if $p_{n+1} = \ep$;
\item
for each $((p_i,\rho_i), l_i, (p_{i+1},\rho_{i+1}))$ with 
$l_i \neq \sfe$, $\Sat{\gD}{[\rho_{i+1},\rho_i](G)}$. 
\end{itemize}

Let $\assproc{R}{G}{\phi}{p}{\psi}$ be an asserted process.
Then $\assproc{R}{G}{\phi}{p}{\psi}$ is 
\emph{true in the sense of partial correctness} iff, 
for all closed substitution instances 
$\assproc{R'}{G'}{\phi'}{p}{\psi'}$ of $\assproc{R}{G}{\phi}{p}{\psi}$,
for all computations $\sigma$ of $p$:
\begin{ldispl}
\sigma \asat (\phi',R') \quad\text{implies}\quad
\sigma \csat (\psi',G')\;.
\end{ldispl}%

In the above definition, the qualification ``in the sense of partial 
correctness'' is used because, if the asserted process is true according 
to this definition, $p$ produces correct results upon termination but 
does not necessarily terminate.

Below, we will present the axioms and rules of \RG.
In addition to axioms and rules that concern a particular constant or 
operator of \deACPei, there is a rule concerning auxiliary 
flexible variables and a rule for pre-condition strengthening, 
rely-condition strengthening, guarantee-condition weakening, and
post-condition weakening.
We use some special notations in the presentation of the axioms and 
rules of \RG.

In the premises of some rules, we write:
\begin{itemize}
\item
$\prevvar{\phi}$, where $\phi \in \stRGCondTerm$, for $\phi$ with, for 
all $v \in \FlexVar$, all occurrences of $v$ replaced by\nolinebreak[2]
$\prevvar{v}$;
\item
$\prevvar{e}$, where $e \in \RGDataTerm$, for $e$ with, for all 
$v \in \FlexVar$, all occurrences of $v$ replaced by\nolinebreak[2]
$\prevvar{v}$;
\item
$\unch(V)$, where $V$ is finite subset of $\FlexVar$, for 
$\LAND_{v \in V} v = \prevvar{v}$.
\end{itemize}

Let $p \in \RGProcTerm$, and let $A \subset \FlexVar$ be such that each 
$v \in A$ occurs in $p$.
Then $A$ is a \emph{set of auxiliary variables of} $p$ if each  
flexible variable from $A$ occurs in $p$ only in subterms of the form 
$\ass{v}{e}$ with $v \in A$.

In the rule concerning auxiliary flexible variables, we write:
\begin{itemize}
\item
$\FVar(\Phi)$, where $\Phi \subset \trRGCondTerm$, for the set of all 
$v \in \FlexVar$ for which $v$ or $\prevvar{v}$ occurs in $\Phi$;
\item
$\AVars(p)$, where $p \in \RGProcTerm$, for the set of all sets of 
auxiliary variables of $p$;
\item
$\phi_A$, where $\phi \in \trRGCondTerm$ and $A \in \AVars(p)$, for a
formula $\Lexists{v_1}{\ldots\Lexists{v_n}{\phi}\ldots}$ such that 
$A = \Set{v_1,\ldots,v_n}$;
\item
$p_A$, where $p \in \RGProcTerm$ and $A \in \AVars(p)$, for $p$ with all 
occurrences of subterms of the form $\ass{v}{e}$ with $v \in A$ replaced 
by $\ep$.
\end{itemize}

The axioms and rules of \RG\ are given in Table~\ref{rules-RG-deACPei}.
\begin{table}[!t]
\caption{Axioms and rules of \RG}
\label{rules-RG-deACPei}
\begin{druletbl}
{} \\[-3ex]
\text{inaction:}
&
\RGAxiom{\assproc{R}{G}{\phi}{\dead}{\psi}}
\\
\text{empty process:}
&
\RGRule
{\Sat{\gD}{\prevvar{\phi} \Land R \Limpl \phi}}
{\assproc{R}{G}{\phi}{\ep}{\phi}}
\\
\text{non-assignment action:}
&
\RGRule
{\begin{array}[t]{@{}c@{}}
 \Sat{\gD}{\prevvar{\phi} \Land R \Limpl \phi},\;\;
 \Sat{\gD}{\prevvar{\phi} \Land \unch(\FVar(\phi,G)) \Limpl G}
 \end{array}}
{\assproc{R}{G}{\phi}{\alpha}{\phi}}
\; \alpha \notin \AProcASS
\\
\text{assignment action:}
&
\RGRule
{\begin{array}[t]{@{}c@{}}
 \Sat{\gD}{\phi \Limpl \psi\subst{e}{v}}, \;\;
 \Sat{\gD}{\prevvar{\phi} \Land R \Limpl \phi}, \;\;
        \Sat{\gD}{\prevvar{\psi} \Land R \Limpl \psi} \\
 \Sat{\gD}{\prevvar{\phi} \Land
           (v = \prevvar{e} \Lor v = \prevvar{v}) \Land
           \unch(\FVar(\phi,G) \Sdiff \Set{v}) \Limpl G} 
 \end{array}}
{\assproc{R}{G}{\phi}{\ass{v}{e}}{\psi}}
\\
\text{alternative composition:}
&
\RGRule
{\begin{array}[t]{@{}c@{}}
 \assproc{R}{G}{\phi}{p}{\psi},\;\; \assproc{R}{G}{\phi}{q}{\psi}
 \end{array}}
{\assproc{R}{G}{\phi}{p \altc q}{\psi}}
\\
\text{sequential composition:}
&
\RGRule
{\assproc{R}{G}{\phi}{p}{\chi},\;\; \assproc{R}{G}{\chi}{q}{\psi}}
{\assproc{R}{G}{\phi}{p \seqc q}{\psi}}
\\
\text{iteration:}
&
\RGRule
{\assproc{R}{G}{\phi}{p}{\phi},\;\; \assproc{R}{G}{\phi}{q}{\psi}}
{\assproc{R}{G}{\phi}{p \iter q}{\psi}}
\\
\text{guarded command:}
&
\RGRule
{\Sat{\gD}{\prevvar{\phi} \Land R \Limpl \phi},\;\;
 \assproc{R}{G}{\phi \Land \chi}{p}{\psi}}
{\assproc{R}{G}{\phi}{\chi \gc p}{\psi}}
\\
\text{parallel composition:}
&
\RGRule
{\assproc{R \Lor G''}{G'}{\phi}{p}{\psi'},\;\;
 \assproc{R \Lor G'}{G''}{\phi}{q}{\psi''}}
{\assproc{R}{G' \Lor G''}{\phi}{p \parc q}{\psi' \Land \psi''}}
\\
\text{encapsulation:}
&
\RGRule
{\assproc{R}{G}{\phi}{p}{\psi}}
{\assproc{R}{G}{\phi}{\encap{H}(p)}{\psi}}
\\
\text{auxiliary variable:}
&
\RGRule
{\begin{array}[t]{@{}c@{}}
 \Sat{\gD}{\prevvar{\phi} \Land R \Limpl \phi},\;\;
 \Sat{\gD}{\phi'_A}, \;\; \Sat{\gD}{R'_A} \\
 \assproc{R \Land R'}{G}{\phi \Land \phi'}{p}{\psi}
 \end{array}}
{\assproc{R}{G}{\phi}{p_A}{\psi}}
\; 
\begin{array}[c]{@{}l@{}}
A \in \AVars(p) \\ 
\FVar(\Set{\phi,\psi,R,G}) \Sinter A = \emptyset
\end{array}
\\
\text{consequence:}
&
\RGRule
{\begin{array}[t]{@{}c@{}}
 \Sat{\gD}{\phi \Limpl \phi'}, \;\; \Sat{\gD}{R \Limpl R'}, \;\;
 \Sat{\gD}{G' \Limpl G},\;\; \Sat{\gD}{\psi' \Limpl \psi} \\
 \assproc{R'}{G'}{\phi'}{p}{\psi'}
 \end{array}}
{\assproc{R}{G}{\phi}{p}{\psi}}
\\[-1.5ex]
\end{druletbl}
\end{table}
In this table, $p$ and~$q$ stand for arbitrary terms from $\RGProcTerm$,\,
$R$, $R'$, $G$, $G'$, and $G''$ stand for arbitrary terms from 
$\trRGCondTerm$,\,
$\phi$, $\phi'$, $\psi$, $\psi'$, $\psi''$, and $\chi$ stand for 
arbitrary terms from $\stRGCondTerm$,\,
$\alpha$ stands for an arbitrary term from $\RGAProcTerm$,\,
$v$ stands for an arbitrary flexible variable from $\FlexVar$,\,
$e$ stands for an arbitrary term from $\DataTerm$,\, and
$A$ stands for an arbitrary subset of $\FlexVar$.

In many rules, not all premises are asserted processes.
The additional premises are judgments of the form $\Sat{\gD}{\phi}$, 
where $\phi$ is a term from $\trRGCondTerm$.
Let $\Sat{\gD}{\phi}$ be such a premise.
Then $\Sat{\gD}{\phi}$ is \emph{true} if $\phi$ holds in $\gD$.

A premise of the form $\Sat{\gD}{\prevvar{\phi} \Land R \Limpl \phi}$,
where $\phi$ is a pre- or post-condition \linebreak[2] and $R$ is a 
rely-condition, expresses that the satisfaction of $\phi$ is preserved 
by steps that satisfy $R$.
In the non-assignment action rule and the assignment action rule, two 
premises of this form occur and in the guarded command rule and the 
auxiliary variable rule one premise of this form occurs. 
Proposition~\ref{proposition-preservation}, given below, tells us that 
premises of this form are unnecessary in rules other than those 
mentioned above.
A premise of the form $\Sat{\gD}{\prevvar{\phi} \Land E \Limpl G}$,
where $\phi$ is a pre-condition, $E$ is a formula representing the 
effect of a step, and $G$ is a guarantee-condition, expresses that, if 
$\phi$ is satisfied immediately before a step with the effect 
represented by $E$, then that step satisfies $G$.
A premise of this form occurs only in the non-assignment action rule and 
the assignment action rule.

The rely-conditions in the premises of the parallel composition rule 
express that in the parallel composition of two processes one process 
can rely on what is relied by the parallel composition or guaranteed by 
the other process.
The guarantee-condition in the conclusion of the parallel composition 
rule express that the parallel composition of two processes can 
guarantee what is guaranteed by either one or the other.

Most rules of \RG\ are simple generalizations of the corresponding rules 
of the Hoare logic of asserted processes presented in~\cite{BM19b}.
The non-assignment action rule and the assignment action rule are 
relatively complex because they have to deal with a pre- and 
post-condition preservation property and a frame problem (flexible 
variables that are not explicitly changed remain unchanged).
The parallel composition rule is relatively complex because it has to 
deal with interference between parallel processes.

Concerning the pre- and post-condition preservation property, we have 
the following result.
\begin{proposition}
\label{proposition-preservation}
Let $\assproc{R}{G}{\phi}{p}{\psi}$ be an asserted process  
derivable from the axioms and rules of \RG.
Then there exist $\phi',\psi' \in \stRGCondTerm$ and 
{$R' \in \trRGCondTerm$} with $\phi \Limpl \phi'$, $R \Limpl R'$, and
$\psi' \Limpl \psi$ such that 
$\Sat{\gD}{\prevvar{\phi'} \Land R' \Limpl \phi'}$ and 
$\Sat{\gD}{\prevvar{\psi'} \Land R' \Limpl \psi'}$.
\end{proposition}
\begin{proof}
This follows directly from the axioms and rules of \RG\ by induction on 
the structure of $p$.
\qed
\end{proof}
This result is lost if the premise
$\Sat{\gD}{\prevvar{\phi} \Land R \Limpl \phi}$ is removed from the 
guarded command rule and/or the auxiliary variable rule.

Although logics like \RG\ are usually called compositional, the use of
the auxiliary variable rule actually breaks compositionality: the 
process referred to in the conclusion of the rule is not composed of the 
processes referred to in the premises.
Therefore, it is discouraged to use auxiliary variables.
There are usually two problems that lead to the use of auxiliary 
variables anyway:
\begin{itemize}
\item
the interference between two parallel processes cannot be fully 
expressed by means of the rely- and guarantee-conditions of asserted
processes for those processes without using auxiliary variables;
\item
an asserted process that is true cannot be derived from the axioms and 
rules of \RG\ without using auxiliary variables.
\end{itemize}
For either problem, it is an open question whether there exists a 
compositional alternative in the setting of \RG\ that solves the 
problem. 
In a closely related setting, the circumvention of the use of auxiliary 
variables in the case of the first problem, by means of concepts called 
``phased specification'' and ``possible values'', is explored 
in~\cite{JP11a,JY19a,Yat24a}.
However, these explorations have not yet led to a more expressive 
compositional logic in rely/guarantee style.
Theorem~5 from \cite{GR16a} suggests that the replacement of the 
auxiliary variable rule by a so-called ``adaptation rule'', which does 
not break compositionality, will not solve the second problem in the 
case of \RG.

The  following result concerning the step relation $\step{}$ is a 
corollary of Lemmas~\ref{lemma-HNF} and~\ref{lemma-summand}.
\begin{corollary}
\label{corollary-step-rel}
For all $p,p' \in \RGProcTerm$, $\rho,\rho' \in \FVarVal$, and 
$\alpha \in \AProcTerm$,
$(p,\rho) \step{\alpha} (p',\rho')$ iff 
there exist $q \in \HNF$ and $q' \in \ProcTerm$ such that 
$p = q$ and $p' = q'$ are derivable from the axioms of \deACPei\ and
$\True \gc \alpha \seqc \eval{\rho'}(q') \sqsubseteq \eval{\rho}(q)$.
\end{corollary}
This corollary shows that, although there is no completeness proof for
the axiom system of \deACPei, this axiom system allows to establish by 
equational reasoning, for all $p,p' \in \RGProcTerm$, 
$\rho,\rho' \in \FVarVal$, and $\alpha \in \AProcTerm$, whether 
$(p,\rho) \step{\alpha} (p',\rho')$ holds.

\section{Soundness of the Axioms and Rules of \RG}
\label{sect-soundness-RG}

This section concerns the soundness of the axioms and rules of \RG\ with
respect to truth in the sense of partial correctness.
It is straightforward to prove that each axiom is true and each of the 
rules, except the parallel composition rule, is such  that only true 
conclusions can be drawn from true premises. 
In this section, the attention is focussed on the parallel composition 
rule.
To prove that this rule is such that only true conclusions can be drawn 
from true premises, it is useful to prove first some lemmas.
In those lemmas an auxiliary notion is used and in their proofs some 
special notation is used.
The special notation and auxiliary notion are introduced first.

Let $\sigma = 
 ((p_1,\rho_1), l_1, (p_2,\rho_2)), \ldots,
 ((p_n,\rho_n), l_n, (p_{n+1},\rho_{n+1}))$
be a computation.
Then define:
\begin{ldispl}
\renewcommand{\arraystretch}{1}
\begin{array}[t]{@{}lll@{}}
\len(\sigma)    & = n;
\\
\proc(\sigma,i) & = p_i    & \;\text{for}\; i = 1, \ldots, n + 1\;,
\\
\val(\sigma,i)  & = \rho_i & \;\text{for}\; i = 1, \ldots, n + 1\;,
\\
\lbl(\sigma,i)  & = l_i    & \;\text{for}\; i = 1, \ldots, n\;.
\end{array}
\end{ldispl}%

Let $p',p'' \in \RGProcTerm$, and 
let $\sigma$ be a computation of $p' \parc p''$, 
$\sigma'$ be a computation of $p'$, and 
$\sigma''$ be a computation of $p''$.
Then $\sigma$, $\sigma'$, and $\sigma''$ \emph{conjoin},
written $\sigma \propto \sigma' \parc \sigma''$, iff
\begin{itemize}
\item
$\len(\sigma) = \len(\sigma') = \len(\sigma'')$;
\item
$\proc(\sigma,i) = \proc(\sigma',i) \parc \proc(\sigma'',i)$
for $i = 1, \ldots, \len(\sigma) + 1$;
\item
$\val(\sigma,i) = \val(\sigma',i) = \val(\sigma'',i)$
for $i = 1, \ldots, \len(\sigma) + 1$;
\item
one of the following:
\begin{itemize}
\item
$\lbl(\sigma',i) \neq \sfe$, $\lbl(\sigma'',i) = \sfe$, and
$\lbl(\sigma,i) = \lbl(\sigma',i)$;
\item
$\lbl(\sigma',i) = \sfe$, $\lbl(\sigma'',i) \neq \sfe$, and
$\lbl(\sigma,i) = \lbl(\sigma'',i)$;
\item
$\lbl(\sigma',i) = \sfe$, $\lbl(\sigma'',i) = \sfe$, and
$\lbl(\sigma,i) = \sfe$;
\item
$\lbl(\sigma',i) \neq \sfe$, $\lbl(\sigma'',i) \neq \sfe$, and
there exists an $\alpha \in \AProcTerm$ such that \\ the equation\,
$\lbl(\sigma',i) \commm \lbl(\sigma'',i) = \alpha$ is derivable and 
$\lbl(\sigma,i) = \alpha$
\end{itemize}
for $i = 1, \ldots, \len(\sigma)$.
\end{itemize}

\begin{lemma}
\label{lemma-conjoin}
Let $p',p'' \in \RGProcTerm$, and 
let $\sigma$ be a computation of some $p \in \RGProcTerm$.
Then $\sigma$ is a computation of $p' \parc p''$ iff
there exist a computation $\sigma'$ of $p'$ and a computation $\sigma''$ 
of $p''$ such that $\sigma \propto \sigma' \parc \sigma''$.
\end{lemma}
\begin{proof}
This follows directly from the definition of the step relation $\step{}$ 
and the structural operational semantics of \deACPei.
\qed
\end{proof}

\begin{lemma}
\label{lemma-soundness-parc-i}
Assume $\assproc{R \Lor G''}{G'}{\phi}{p}{\psi'}$ and
$\assproc{R \Lor G'}{G''}{\phi}{q}{\psi''}$ are true.
Let $\sigma$ be a computation of $p \parc q$ such that
$\sigma \asat(\phi,R)$, and
let $\sigma'$ be a computation of $p$ and $\sigma''$ be a computation of 
$q$ such that $\sigma \propto \sigma' \parc \sigma''$.
Then:
\begin{enumerate}
\item[\textup{1.}]
for $i = 1, \ldots, \len(\sigma)$:
\begin{enumerate}  
\item[\textup{(a)}]
$\Sat{\gD}{[\val(\sigma',i),\val(\sigma',i+1)](G')}$\,\,\,\, 
if $\lbl(\sigma',i) \neq \sfe$;
\item[\textup{(b)}]
$\Sat{\gD}{[\val(\sigma'',i),\val(\sigma'',i+1)](G'')}$ 
if $\lbl(\sigma'',i) \neq \sfe$;
\end{enumerate}  
\item[\textup{2.}]
for $i = 1, \ldots, \len(\sigma)$:
\begin{enumerate}  
\item[\textup{(a)}]
$\Sat{\gD}{[\val(\sigma',i),\val(\sigma',i+1)](R \Lor G'')}$\, 
if $\lbl(\sigma',i) = \sfe$;
\item[\textup{(b)}]
$\Sat{\gD}{[\val(\sigma'',i),\val(\sigma'',i+1)](R \Lor G')}$ 
if $\lbl(\sigma'',i) = \sfe$.
\end{enumerate}  
\end{enumerate}
\end{lemma}
\begin{proof}
\mbox{} \\
Part~1.\,
Assume that Part~1 is not the case.
Then two cases can be distinguished:
\begin{itemize}
\item
there exists a prefix $\varsigma'$ of $\sigma'$ such that:
\begin{center}
\renewcommand{\arraystretch}{1.25}
\begin{tabular}[t]{@{}l@{}}
$\nSat{\gD}
      {[\val(\varsigma',\len(\varsigma')),
        \val(\varsigma',\len(\varsigma')+1)](G')}$ and
$\lbl(\varsigma',\len(\varsigma')) \neq \sfe$;
\\
$\Sat{\gD}{[\val(\sigma'',i),\val(\sigma'',i+1)](G'')}$
if $\lbl(\sigma'',i)) \neq \sfe$ for $i = 1, \ldots, \len(\varsigma')$;
\end{tabular}
\end{center}
\item
there exists a prefix $\varsigma''$ of $\sigma''$ such that:
\begin{center}
\renewcommand{\arraystretch}{1.25}
\begin{tabular}[t]{@{}l@{}}
$\nSat{\gD}
      {[\val(\varsigma'',\len(\varsigma'')),
        \val(\varsigma'',\len(\varsigma'')+1)](G'')}$ and
$\lbl(\varsigma'',\len(\varsigma'')) \neq \sfe$;
\\ 
$\Sat{\gD}{[\val(\sigma',i),\val(\sigma',i+1)](G')}$
if $\lbl(\sigma',i)) \neq \sfe$ for $i = 1, \ldots, \len(\varsigma'')$.
\end{tabular}
\end{center}
\end{itemize}
Because the two cases are symmetric, it will only be shown that the 
first case leads to a contradiction.

Let $\varsigma'$ be the shortest prefix of $\sigma'$ such that:
\begin{center}
\renewcommand{\arraystretch}{1.25}
\begin{tabular}[t]{@{}l@{}}
$\nSat{\gD}
      {[\val(\varsigma',\len(\varsigma')),
        \val(\varsigma',\len(\varsigma')+1)](G')}$ and
$\lbl(\varsigma',\len(\varsigma')) \neq \sfe$;
\\
$\Sat{\gD}{[\val(\sigma'',i),\val(\sigma'',i+1)](G'')}$
if $\lbl(\sigma'',i)) \neq \sfe$ for $i = 1, \ldots, \len(\varsigma')$.
\end{tabular}
\end{center}
It follows from Lemma~\ref{lemma-conjoin} that, for 
$i = 1, \ldots, \len(\varsigma')$, if $\lbl(\varsigma',i) = \sfe$ then
either $\lbl(\sigma'',i) \neq \sfe$ or $\lbl(\sigma,i) = \sfe$.
This means that $\varsigma' \asat(\phi,R \Lor G'')$.
However,  
because 
$\nSat{\gD}
      {[\val(\varsigma',\len(\varsigma')),
        \val(\varsigma',\len(\varsigma')+1)](G')}$
and $\lbl(\varsigma',\len(\varsigma')) \neq \sfe$,
this contradicts $\assproc{R \Lor G''}{G'}{\phi}{p}{\psi'}$.
Hence, Part~1 is the case.
\\[1ex]
Part~2.\, 
It follows from Lemma~\ref{lemma-conjoin} that, for 
$i = 1, \ldots, \len(\sigma')$, if $\lbl(\sigma',i) = \sfe$ then
either $\lbl(\sigma'',i) \neq \sfe$ or $\lbl(\sigma,i) = \sfe$.
Moreover, it follows from Part~1 that, 
for $i = 1, \ldots, \len(\sigma')$,
$\Sat{\gD}{[\val(\sigma'',i),\val(\sigma'',i+1)](G')}$ 
if $\lbl(\sigma'',i) \neq \sfe$.
Together this means that, for $i = 1, \ldots, \len(\sigma')$, 
$\Sat{\gD}{[\val(\sigma',i),\val(\sigma',i+1)](R \Lor G'')}$.
This proves~(a). The proof of~(b) goes analogously.
\qed
\end{proof}

\begin{lemma}
\label{lemma-soundness-parc-ii}
Assume $\assproc{R \Lor G''}{G'}{\phi}{p}{\psi'}$ and
$\assproc{R \Lor G'}{G''}{\phi}{q}{\psi''}$ are true.
Let $\sigma$ be a computation of $p \parc q$ such that
$\sigma \asat(\phi,R)$, and
let $\sigma'$ be a computation of $p$ and $\sigma''$ be a computation of 
$q$ such that $\sigma \propto \sigma' \parc \sigma''$.
Then:
\begin{enumerate}
\item[\textup{1.}]
for $i = 1, \ldots, \len(\sigma)$: 
\\ \hspace*{1.25em}
$\Sat{\gD}{[\val(\sigma,i),\val(\sigma,i+1)](G' \Lor G'')}$ 
if $\lbl(\sigma,i) \neq \sfe$;
\item[\textup{2.}]
$\Sat{\gD}{\val(\sigma,\len(\sigma)+1)(\psi' \Land \psi'')}$ 
if $\proc(\sigma,\len(\sigma)+1) = \ep$.
\end{enumerate}
\end{lemma}
\begin{proof}
\mbox{} \\
Part~1.\,
It follows from Lemma~\ref{lemma-conjoin} that, for 
$i = 1, \ldots, \len(\sigma)$, if $\lbl(\sigma,i) \neq \sfe$ then
either $\lbl(\sigma',i) \neq \sfe$ or $\lbl(\sigma'',i) \neq \sfe$.
From this and Part~1 of Lemma~\ref{lemma-soundness-parc-i} it follows 
that, for $i = 1, \ldots, \len(\sigma)$, 
$\Sat{\gD}{[\val(\sigma,i),\val(\sigma,i+1)](G' \Lor G'')}$ if 
$\lbl(\sigma,i) \neq \sfe$.
\\[1ex]
Part~2.\,
Assume that $\proc(\sigma,\len(\sigma)+1) = \ep$.
Then, by the definition of $\sigma \propto \sigma' \parc \sigma''$,
$\proc(\sigma',\len(\sigma)+1) = \ep$ and
$\proc(\sigma'',\len(\sigma)+1) = \ep$.
Moreover, it follows from Part~2 of Lemma~\ref{lemma-soundness-parc-i} 
that $\sigma' \asat(\phi,R \Lor G'')$ and 
$\sigma'' \asat(\phi,R \Lor G')$.
Together this means that 
$\Sat{\gD}{\val(\sigma',\len(\sigma)+1)(\psi')}$ and
$\Sat{\gD}{\val(\sigma'',\len(\sigma)+1)(\psi'')}$.
From this, it follows, by the definition of 
$\sigma \propto \sigma' \parc \sigma''$, that
$\Sat{\gD}{\val(\sigma,\len(\sigma)+1)\linebreak[2](\psi' \Land \psi'')}$.
\qed
\end{proof}

\begin{theorem}[Soundness]
\label{theorem-soundness-RG}
For all $p \in \RGProcTerm$, $\phi,\psi \in \stRGCondTerm$, and 
$R,G \in \trRGCondTerm$, the asserted process 
$\assproc{R}{G}{\phi}{p}{\psi}$ is derivable from the axioms and rules 
of \RG\ only if $\assproc{R}{G}{\phi}{p}{\psi}$ is true in the sense of 
partial correctness.
\end{theorem}
\begin{proof} 
By the definition of the truth of asserted processes, it is sufficient 
to consider only $\phi$, $\psi$, $R$, and $G$ that are closed terms.
The theorem is proved by proving that each of the axioms is true and
each of the rules is such that only true conclusions can be drawn from 
true premises.   
The theorem then follows by induction on the length of the proof.

It is straightforward to prove that each of the axioms is true and
each of the rules, except the parallel composition rule, is such that 
only true conclusions can be drawn from true premises.   
That the parallel composition rule is also such that only true 
conclusions can be drawn from true premises follows immediately from 
Parts~1 and~2 of Lemma~\ref{lemma-soundness-parc-ii}.
\qed
\end{proof}

\section{A Simple Example}
\label{sect-example}

Below, we describe the behaviour of a very simple system by a closed 
\deACPei\ \linebreak[2]term and reason about how this system changes 
data with the axioms and rules of \RG.
We assume that $\gD$'s carrier of sort $\Data$ is the set of all 
integers, that $\sign_\gD$ includes all constants and operators used 
below, and that the interpretation of these constants and operators in
$\gD$ is as usual.
We also assume that $i$ and $j$ are flexible variables from $\FlexVar$ 
and $n$ and $n'$ are variables of sort $\Data$.
The behaviour of the very simple system concerned is described by the 
closed \deACPei\ term 
$\ass{i}{i+1} \seqc \ass{i}{i+1} \parc \ass{i}{0}$. 

In \RG, the asserted process
\begin{ldispl}
\langle i = \prevvar{i},i = \prevvar{i} + 1 \Lor i = 0 \rangle : 
\\ \qquad
\hlassproc{i = 0}{\ass{i}{i+1} \seqc \ass{i}{i+1} \parc \ass{i}{0}}
  {i = 0 \Lor i = 1 \Lor i = 2}
\end{ldispl}%
is true in the sense of partial correctness.
Below, it is shown how this asserted process can be derived by means of 
the axioms and rules of \RG.

We derive, using the assignment action rule, the following two asserted
processes:
\begin{ldispl}
\langle i = 0 \Lor i = \prevvar{i},
        i = \prevvar{i} + 1 \Lor i = \prevvar{i}\, \rangle :
\\ \qquad
\hlassproc{i = 0}{\ass{i}{i+1}}{i = 0 \Lor i = 1}\;,
\eqnsep
\langle i = 0 \Lor i = \prevvar{i},
        i = \prevvar{i} + 1 \Lor i = \prevvar{i}\, \rangle :
\\ \qquad
\hlassproc{i = 0 \Lor i = 1}{\ass{i}{i+1}}
  {i = 0 \Lor i = 1 \Lor i = 2}\;.
\end{ldispl}%
From these two asserted processes, we derive, using the sequential 
composition rule, the following asserted process:
\begin{ldispl}
\langle i = 0 \Lor i = \prevvar{i},
        i = \prevvar{i} + 1 \Lor i = \prevvar{i}\, \rangle :
\\ \qquad
\hlassproc{i = 0}{\ass{i}{i+1} \seqc \ass{i}{i+1}}
  {i = 0 \Lor i = 1 \Lor i = 2}\;.
\end{ldispl}%
We derive, also using the assignment action rule, the following asserted
process:
\begin{ldispl}
\assproc{i = \prevvar{i} + 1 \Lor i = \prevvar{i}}
  {i = 0 \Lor i = \prevvar{i}\,}
  {i = 0}{\ass{i}{0}}{\True}\;.
\end{ldispl}%
From the last two asserted processes, we derive, using the parallel 
composition rule, the following asserted process:
\begin{ldispl}
\langle i = \prevvar{i},i = \prevvar{i} + 1 \Lor i = 0 \rangle : 
\\ \qquad
\hlassproc{i = 0}{\ass{i}{i+1} \seqc \ass{i}{i+1} \parc \ass{i}{0}}
  {i = 0 \Lor i = 1 \Lor i = 2}\;.
\end{ldispl}%

From the last asserted process, we further derive, using the consequence 
rule, an asserted process that concerns the process 
$\ass{i}{i+1} \seqc \ass{i}{i+1} \parc \ass{i}{0}$ 
considered in isolation:
\begin{ldispl}
\assproc{\False}{\True}
  {i = 0}{\ass{i}{i+1} \seqc \ass{i}{i+1} \parc \ass{i}{0}}
  {i = 0 \Lor i = 1 \Lor i = 2}\;.
\end{ldispl}%

In~\cite{BM19b}, a Hoare logic for \deACPei\ is presented.
It is not hard to see that for each asserted process 
$\hlassproc{\phi}{p}{\psi}$ from that Hoare logic:
\begin{ldispl}
\hlassproc{\phi}{p}{\psi} 
\;\text{is true in the sense of partial correctness} \\
\quad\text{iff}\quad
\assproc{\False}{\True}{\phi}{p}{\psi} 
\;\text{is true in the sense of partial correctness}. 
\end{ldispl}%
This means that, in that Hoare logic, the asserted process
\begin{ldispl}
\hlassproc{i = 0}{\ass{i}{i+1} \seqc \ass{i}{i+1} \parc \ass{i}{0}}
  {i = 0 \Lor i = 1 \Lor i = 2} 
\end{ldispl}%
is true in the sense of partial correctness.
However, this asserted process cannot be derived by means of the axioms 
and rules of the Hoare logic alone because of the rather restrictive 
side condition of its parallel composition rule.
In~\cite{BM19b}, it is shown how reasoning with the axioms and rules of 
the Hoare logic can be combined with equational reasoning with the 
axioms of \deACPei\ to prove the truth of this asserted process. 
It involves 15~applications of axioms and rules of the Hoare logic 
from~\cite{BM19b} after tens of applications of axioms of 
\mbox{\deACPei}.  
As shown above, the derivation of the corresponding asserted process 
from \RG\ involves 6~applications of rules of \RG\ and no applications
of axioms of \deACPei. 

\RG\ does not only prevent equational reasoning with the axioms of 
\deACPei, but can also potentially reduce the required number of 
applications of axioms and rules exponentially.
This follows immediately from the fact that the number of occurrences of 
constants and operators in a term from $\ProcTerm$ can potentially grow 
exponentially by the elimination of the occurrences of the operators 
$\parc$, $\leftm$, and $\commm$ from the term.

\section{Weak Total Correctness}
\label{sect-convergence}

\RG\ is a rely/guarantee logic for partial correctness.
In this section, it is outlined how it can be turned into a 
rely/guarantee logic for weak total correctness.

Informally, an asserted process $\assproc{R}{G}{\phi}{p}{\psi}$ is true 
in the sense of weak total correctness if the following is the case:
\begin{quote}
if $\phi$ holds at the start of $p$ and $R$ holds for each step taken by
processes taking place in parallel with $p$, then 
$G$ holds for each step taken by $p$, $p$~does not keep taking steps into 
infinity, and, if $p$ eventually terminates successfully, $\psi$ holds 
upon successful termination of $p$. 
\end{quote}

The adaptation of \RG\ to weak total correctness requires to consider 
both finite and infinite computations.

Let $p \in \RGProcTerm$, and let
$\sigma = 
 ((p_1,\rho_1), l_1, (p_2,\rho_2)), 
 ((p_2,\rho_2), l_2, (p_3,\rho_3)), \ldots$
be a finite or infinite computation of $p$.
Then $\sigma$ \emph{is convergent}, written $\conv \sigma$, iff 
the set $\Set{l_i \in \Set{l_1,l_2,\ldots} \where l_i \neq \sfe}$ is 
finite.

As for $\gD$, the adaptation to weak total correctness requires 
additional assumptions:
\begin{itemize}
\item
the signature $\sign_\gD$ includes:
\begin{itemize}
\item
a binary operator $\funct{<}{\Data \Sx \Data}{\Bool}$;
\item
a sort $\Ord$ of \emph{ordinals};
\item
constants $\const{0}{\Ord}$ and $\const{\omega}{\Ord}$;
\item
a unary operator $\funct{\ord}{\Data}{\Ord}$;
\item
a binary operator $\funct{<}{\Ord \Sx \Ord}{\Bool}$;
\end{itemize}
\item
the algebra $\gD$ is such that:
\begin{itemize}
\item
the interpretation of $\funct{<}{\Data \Sx \Data}{\Bool}$ is the
characteristic function of a strict well-order on $\Data^\gD$;
\item
$\Ord^\gD$ is the set of all ordinals $\alpha$ such that
not $\omega < \alpha$ (here $<$ denotes the strict order on ordinals and 
$\omega$ denotes the smallest limit ordinal);
\item
the interpretation of $\const{0}{\Ord}$ is the smallest ordinal;
\item
the interpretation of $\const{\omega}{\Ord}$ is the smallest limit 
ordinal;
\item
the interpretation of $\funct{<}{\Ord \Sx \Ord}{\Bool}$ is the
characteristic function of the strict order of ordinals restricted to
$\Ord^\gD$;
\item
the interpretation of $\funct{\ord}{\Data}{\Ord}$ is the unique function 
$\mathit{ord}$ from $\Data^\gD$ to~$\Ord^\gD$ such that the 
corestriction of $\mathit{ord}$ onto the image of $\Data^\gD$ under 
$\mathit{ord}$ is an order isomorphism.
\end{itemize}
\end{itemize}
In addition, the following operator must be added to the constants and 
operators to build term of sort $\Cond$:
\begin{itemize}
\item
a binary \emph{equality} operator $\funct{=}{\Ord \Sx \Ord}{\Cond}$.
\end{itemize}

Because of the addition of sort $\Ord$, it is also assumed that there 
is a countably infinite set of variables of sort $\Ord$ that is disjoint
from the sets of variables of sort $\Proc$, $\Cond$, $\Data$, and the 
set $\FlexVar$.

The additional assumptions concerning $\gD$ are such that the following 
formulas hold in $\gD$:
\begin{ldispl}
\Lforall{X}{\Lforall{Y}{\ord(X) = \ord(Y) \;\Liff\; X = Y}}\;, 
\quad
\Lforall{\alpha}
 {\Lexists{X}{\ord(X) = \alpha} \Liff \Lnot\, \alpha = \omega}\;,
\\
\Lforall{X}{\Lforall{Y}{\ord(X) < \ord(Y) \;\Liff\; X < Y}}\;, 
\quad
\Lforall{X}{\ord(X) < \omega}\;.
\end{ldispl}%
Why the smallest limit ordinal $\omega$ is included in $\Ord^\gD$ is 
explained after the iteration rule adapted to weak total correctness has 
been presented.

Let $\assproc{R}{G}{\phi}{p}{\psi}$ be an asserted process.
Then $\assproc{R}{G}{\phi}{p}{\psi}$ is 
\emph{true in the sense of weak total correctness} iff, 
for all closed substitution instances 
$\assproc{R'}{G'}{\phi'}{p}{\psi'}$ of $\assproc{R}{G}{\phi}{p}{\psi}$,
for all computations $\sigma$ of $p$:
\begin{ldispl}
\sigma \asat (\phi',R') \quad\text{implies}\quad
\conv \sigma \;\;\text{and}\;\; \sigma \csat (\psi',G')\;.
\end{ldispl}%

The axioms and rules of the rely/guarantee logic has to be adapted to
weak total correcness.
Only the iteration rule needs to be changed.
It becomes:
\begin{ldispl}
\RGRule
{\begin{array}[c]{@{}c@{}}
 \Sat{\gD}
  {\prevvar{\phi} \Land R \Limpl
   \Lexists{\alpha'}
    {\phi\subst{\alpha'}{\alpha} \Land \alpha' \leq \alpha}} \\
 \assproc{R}{G}
  {\phi \Land \alpha > 0}{p}
  {\Lexists{\alpha'}
    {\phi\subst{\alpha'}{\alpha} \Land \alpha' < \alpha}}, \;\; 
 \assproc{R}{G}{\phi\subst{0}{\alpha}}{q}{\psi}
 \end{array}}
{\assproc{R}{G}
  {\Lexists{\alpha'}{\phi\subst{\alpha'}{\alpha}}}{p \iter q}{\psi}}
\end{ldispl}%
where $\alpha$ and $\alpha'$ stand for arbitrary variables of sort 
$\Ord$ and $\phi$ stands for an arbitrary term from $\stRGCondTerm$ in 
which $\alpha$ occurs.

The set $\Ord^\gD \Sdiff \Set{\omega}$ is the set of all natural 
numbers.
As for the inclusion of the value $\omega$ in $\Ord^\gD$, consider an 
asserted process of the form 
$\assproc{R}{G}{\phi}{p \iter q}{\psi}$.
A term of the form $\ord(e)$, where $e$ is a term from $\RGDataTerm$,
may be used in $\phi$ to provide an upper bound on the number of 
remaining iterations of $p$.
Due to steps made by the environment, such a term may have any value 
from $\Ord^\gD \Sdiff \Set{\omega}$ immediately before $p$ takes its 
first step.
Because the upper bound on the number of remaining iterations of $p$ 
must be smaller after each iteration, a term may be needed that provides
a value that is greater than all those values.
For this reason, the value $\omega$ is included in $\Ord^\gD$ and the 
constant $\omega$ is included in $\sign_\gD$.
Below is a simple example where $\omega$ is actually needed.

Let $p$ be 
\begin{ldispl}
((i = 0 \Lor j > 0) \gc ((i = 0 \gc \ass{i}{1}) \altc
 (\Lnot\, i = 0 \gc \ass{j}{j - 1})))
\phantom{\;.} \quad \\ \hfill {}
\iter
(\Lnot (i = 0 \Lor j > 0) \gc \ep)\;.
\end{ldispl}%
The asserted process
\begin{ldispl}
\assproc{i = \prevvar{i} \Land
         (\Lnot\, \prevvar{i} = 0 \Limpl j = \prevvar{j})}{\True}
        {i = 0}{p}{\True}
\end{ldispl}%
expresses that,
if the value of $i$ equals $0$ at the start of $p$ and each step taken 
by processes taking place in parallel with $p$ does not change the value 
of $i$ and does not change the value of $j$ if the value of $i$ does not
equal $0$, then $p$ eventually terminates successfully if $p$ is 
deadlock free.

The last two steps of a derivation of this asserted process are an 
application of the adapted iteration rule and an application of the 
consequence rule.
This means that the condition $\phi$ from the adapted iteration rule 
must be such that
\begin{ldispl}
i = 0 \Limpl \Lexists{\alpha'}{\phi\subst{\alpha'}{\alpha}}
\end{ldispl}%
holds in $\gD$.
A derivation of this asserted process involves two applications of a 
rule without asserted processes among its premises, viz.\ the 
assignment action rule.
By consulting the other rules to be applied in a derivation, it is 
easy to see that the condition $\phi$ must also be such that the 
following asserted processes must be derivable by the applications of 
the assignment action rule:
\begin{ldispl}
\langle i = \prevvar{i} \Land
         (\Lnot\, \prevvar{i} = 0 \Limpl j = \prevvar{j}),\True\, 
\rangle : \\ \qquad
\hlassproc{\phi \Land \alpha > 0 \Land i = 0}{\ass{i}{1}}
 {\Lexists{\alpha'}
   {\phi\subst{\alpha'}{\alpha} \Land \alpha' < \alpha}}\;, 
\eqnsep 
\langle i = \prevvar{i} \Land
         (\Lnot\, \prevvar{i} = 0 \Limpl j = \prevvar{j}),\True\,
\rangle : \\ \qquad
\hlassproc{\phi \Land \alpha > 0 \Land \Lnot\, i = 0}{\ass{j}{j - 1}}
 {\Lexists{\alpha'}
   {\phi\subst{\alpha'}{\alpha} \Land \alpha' < \alpha}}\;.
\end{ldispl}%
It follows from the above analysis that the following condition can be 
taken as~$\phi$: 
\begin{ldispl}
(i = 0 \Limpl \alpha = \omega) \Land
(\Lnot\, i = 0 \Limpl \alpha = \ord(j))\;.
\end{ldispl}%
If the constant $\omega$ was not included in $\sign_\gD$, there would be 
no condition that can be taken as $\phi$.

\section{Deadlock Freedom}
\label{sect-deadlock-freedom}

This section concerns the adaptation of the rely/guarantee logic for 
weak total correctness from Section~\ref{sect-convergence} to deadlock 
freedom.

The adaptation to deadlock freedom requires first of all the extension
of asserted processes with an enabledness-condition.

An \emph{asserted process with enabledness-condition} 
is a judgment of the form $\rassproc{R}{\theta}{G}{\phi}{p}{\psi}$, 
where $\assproc{R}{G}{\phi}{p}{\psi}$ is an asserted process and
$\theta \in \stRGCondTerm$.  
Here, $\theta$ is called the \emph{enabledness-condition} of the 
asserted process.

Informally, an asserted process with enabledness-condition 
$\rassproc{R}{\theta}{G}{\phi}{p}{\psi}$ is true in the sense of weak 
total correctness if the following is the case:
\begin{quote}
if $\phi$ holds at the start of $p$ and $R$ holds for each step taken by
processes taking place in parallel with $p$, then $G$ holds for each 
step taken by~$p$, $p$~does not keep taking steps into infinity, $p$ is 
enabled wherever $\theta$ holds, and, if $p$ eventually terminates 
successfully, $\psi$ holds upon successful termination of $p$. 
\end{quote}
In the case that $\rassproc{R}{\True}{G}{\phi}{p}{\psi}$ is true, $p$ is 
deadlock-free and will eventually terminate successfully if $\phi$ holds 
at the start of $p$ and $R$ holds for each step taken by processes 
taking place in parallel with $p$.
This means that truth in the sense of weak total correctness of asserted
processes with enabledness condition in a way covers truth in a sense 
that could be called strong total correctness.

Let $p \in \RGProcTerm$ and $\theta \in \stRGCondTerm$, and let
$\sigma = 
 ((p_1,\rho_1), l_1, (p_2,\rho_2)), \ldots, 
 ((p_n,\rho_n), l_n, \linebreak[2] (p_{n+1},\rho_{n+1}))$
be a computation of $p$.
Then $\sigma$ \emph{satisfies the enabledness-condition~$\theta$}, 
written $\sigma \esat \theta$, iff
\begin{itemize}
\item
if $\Sat{\gD}{\rho_{n+1}(\theta)}$, then $p_{n+1} = \ep$ or there exists 
a $p' \in \RGProcTerm$, $\rho' \in \FVarVal$, and $\alpha \in \AProcTerm$ 
such that $(p_{n+1},\rho_{n+1}) \step{\alpha} (p',\rho')$.
\end{itemize}

Let $\rassproc{R}{\theta}{G}{\phi}{p}{\psi}$ be an asserted process with 
enabledness-condition.
Then $\rassproc{R}{\theta}{G}{\phi}{p}{\psi}$ is 
\emph{true in the sense of weak total correctness} iff, 
for all closed substitution instances 
$\rassproc{R'}{\theta'}{G'}{\phi'}{p}{\psi'}$ of 
$\rassproc{R}{\theta}{G}{\phi}{p}{\psi}$,
for all computations $\sigma$ of $p$:
\begin{ldispl}
\sigma \asat (\phi',R') \quad\text{implies}\quad
\conv \sigma \;\;\text{and}\;\; \sigma \esat \theta' \;\;\text{and}\;\;
 \sigma \csat (\psi',G')\;.
\end{ldispl}%

The axioms and rules of the rely/guarantee logic has to be adapted to
deadlock freedom.

The adapted axioms and rules of \RG\ are given in 
Table~\ref{rules-RGdf-deACPei}.
\begin{table}[!t]
\caption{Adapted axioms and rules of \RG}
\label{rules-RGdf-deACPei}
\begin{druletbl}
{} \\[-3ex]
\text{inaction:}
&
\RGAxiom{\rassproc{R}{\False}{G}{\phi}{\dead}{\psi}}
\\
\text{empty process:}
&
\RGRule
{\Sat{\gD}{\prevvar{\phi} \Land R \Limpl \phi}}
{\rassproc{R}{\theta}{G}{\phi}{\ep}{\phi}}
\\
\text{non-assignment action:}
&
\RGRule
{\begin{array}[t]{@{}c@{}}
 \Sat{\gD}{\prevvar{\phi} \Land R \Limpl \phi}, \;\;
 \Sat{\gD}{\prevvar{\phi} \Land \unch(\FVar(\phi,G)) \Limpl G}
 \end{array}}
{\rassproc{R}{\theta}{G}{\phi}{\alpha}{\phi}}
\; \alpha \notin \AProcASS
\\
\text{assignment action:}
&
\RGRule
{\begin{array}[t]{@{}c@{}}
 \Sat{\gD}{\phi \Limpl \psi\subst{e}{v}}, \;\;
 \Sat{\gD}{\prevvar{\phi} \Land R \Limpl \phi}, \;\;
        \Sat{\gD}{\prevvar{\psi} \Land R \Limpl \psi} \\
 \Sat{\gD}{\prevvar{\phi} \Land
           (v = \prevvar{e} \Lor v = \prevvar{v}) \Land
           \unch(\FVar(\phi,G) \Sdiff \Set{v}) \Limpl G} \\
 \end{array}}
{\rassproc{R}{\theta}{G}{\phi}{\ass{v}{e}}{\psi}}
\\
\text{alternative composition:}
&
\RGRule
{\begin{array}[t]{@{}c@{}}
 \rassproc{R}{\theta'}{G}{\phi}{p}{\psi},\;\;
 \rassproc{R}{\theta''}{G}{\phi}{q}{\psi}
 \end{array}}
{\rassproc{R}{\theta' \Lor \theta''}{G}{\phi}{p \altc q}{\psi}}
\\
\text{sequential composition:}
&
\RGRule
{\rassproc{R}{\theta}{G}{\phi}{p}{\chi},\;\;
 \rassproc{R}{\theta}{G}{\chi}{q}{\psi}}
{\rassproc{R}{\theta}{G}{\phi}{p \seqc q}{\psi}}
\\
\text{iteration:}
&
\RGRule
{\begin{array}[t]{@{}c@{}}
 \Sat{\gD}
  {\prevvar{\phi} \Land R \Limpl
   \Lexists{\alpha'}
    {\phi\subst{\alpha'}{\alpha} \Land \alpha' \leq \alpha}} \\
 \rassproc{R}{\theta'}{G}
  {\phi \Land \alpha > 0}{p}
  {\Lexists{\alpha'}
    {\phi\subst{\alpha'}{\alpha} \Land \alpha' < \alpha}} \\ 
 \rassproc{R}{\theta''}{G}{\phi\subst{0}{\alpha}}{q}{\psi}
 \end{array}}
{\rassproc{R}{\theta' \Lor \theta''}{G}
  {\Lexists{\alpha'}{\phi\subst{\alpha'}{\alpha}}}{p \iter q}{\psi}}
\\
\text{guarded command:}
&
\RGRule
{\Sat{\gD}{\prevvar{\phi} \Land R \Limpl \phi},\;\;
 \rassproc{R}{\True}{G}{\phi \Land \chi}{p}{\psi}}
{\rassproc{R}{\phi \Limpl \chi}{G}{\phi}{\chi \gc p}{\psi}}
\\
\text{parallel composition:}
&
\RGRule
{\begin{array}[t]{@{}c@{}}
 \Sat{\gD}{\psi'' \Limpl \theta'}, \;\;
 \Sat{\gD}{\psi' \Limpl \theta''}, \;\;
 \Sat{\gD}{\theta' \Lor \theta''} \\
 \rassproc{R \Lor G''}{\theta \Land \theta'}{G'}{\phi}{p}{\psi'}\, \\
 \rassproc{R \Lor G'}{\theta \Land \theta''}{G''}{\phi}{q}{\psi''}
\end{array}}
{\rassproc{R}{\theta}{G' \Lor G''}{\phi}{p \parc q}{\psi' \Land \psi''}}
\\
\text{encapsulation:}
&
\RGRule
{\rassproc{R}{\theta}{G}{\phi}{p}{\psi}}
{\rassproc{R}{\theta}{G}{\phi}{\encap{H}(p)}{\psi}}
\; \AProc(p) \Sinter H = \emptyset
\\
\text{auxiliary variable:}
&
\RGRule
{\begin{array}[t]{@{}c@{}}
 \Sat{\gD}{\prevvar{\phi} \Land R \Limpl \phi} \\
 \Sat{\gD}{\phi'_A}, \;\; \Sat{\gD}{R'_A}, \;\; \Sat{\gD}{\theta'_A} \\
 \rassproc{R \Land R'}{\theta \Land \theta'}{G}{\phi \Land \phi'}{p}{\psi}
 \end{array}}
{\rassproc{R}{\theta}{G}{\phi}{p_A}{\psi}}
\; 
\begin{array}[c]{@{}l@{}}
A \in \AVars(p) \\ 
\FVar(\Set{\phi,\psi,\theta,R,G}) \Sinter A = \emptyset
\end{array}
\\
\text{consequence:}
&
\RGRule
{\begin{array}[t]{@{}c@{}}
 \Sat{\gD}{\phi \Limpl \phi'}, \;\; \Sat{\gD}{R \Limpl R'}, \;\;
 \Sat{\gD}{\theta \Limpl \theta'} \\
 \Sat{\gD}{G' \Limpl G}, \;\; \Sat{\gD}{\psi' \Limpl \psi} \\
 \rassproc{R'}{\theta'}{G'}{\phi'}{p}{\psi'}
 \end{array}}
{\rassproc{R}{\theta}{G}{\phi}{p}{\psi}}
\\[-1.5ex]
\end{druletbl}
\end{table}
In this table, $p$ and~$q$ stand for arbitrary terms from $\RGProcTerm$,\,
$R$, $R'$, $G$, $G'$, and $G''$ stand for arbitrary terms from 
$\trRGCondTerm$,\,
$\phi$, $\phi'$, $\psi$, $\psi'$, $\psi''$, $\theta$, $\theta'$, and 
$\chi$ stand for arbitrary terms from $\stRGCondTerm$,\,
$\alpha$ stands for an arbitrary term from $\RGAProcTerm$,\,
$v$ stands for an arbitrary flexible variable from $\FlexVar$,\,
$e$ stands for an arbitrary term from $\DataTerm$,\, and
$A$ stands for an arbitrary subset of $\FlexVar$.
In the adapted encapsulation rule, we write $\AProc(p)$, where 
$p \in \RGProcTerm$, for the set of all $\alpha \in \AProcTerm$ that 
occur in $p$.

The adapted encapsulation rule is applicable only if $p = \encap{H}(p)$.
This means that this rule has an extremely restrictive side condition.
The following paragraph explains why a better adapted encapsulation rule 
is not feasible.

Assume that $\rassproc{R}{\theta}{G}{\phi}{p}{\psi}$ is an asserted 
process (with enabledness-condition) that is true.
Let 
$((p_1,\rho_1), l_1, (p_2,\rho_2)), \ldots, 
 ((p_n,\rho_n), l_n, (p_{n+1},\rho_{n+1}))$ 
be a computation of $p$ such that 
there exists an $\alpha \in H$ such that, 
for all \mbox{$\rho \in \FVarVal$},
\smash{$\astep{\eval{\rho_{n+1}}(p_{n+1})}{\gact{\rho}
         {\alpha}}{\eval{\rho'}(p')}$}
for some $p'$ and $\rho'$.
Moreover, \mbox{suppose} that the above computation is such that
$((\encap{H}(p_1),\rho_1), l_1, (\encap{H}(p_2),\rho_2)), \ldots,
 \linebreak[2]
 ((\encap{H}(p_n),\rho_n), l_n, (\encap{H}(p_{n+1}),\rho_{n+1}))$
is a computation of $\encap{H}(p)$.
Then
not for all $\rho \in \FVarVal$,
\smash{$\astep{\eval{\rho_{n+1}}(\encap{H}(p_{n+1}))}{\gact{\rho}
         {\alpha}}{\eval{\rho'}(p')}$}
for some $p'$ and $\rho'$.  
However, from the truth of 
$\rassproc{R}{\theta}{G}{\phi}{p}{\psi}$, 
it cannot be determined whether there exists an $\alpha' \notin H$ such 
that 
for all $\rho \in \FVarVal$,
\smash{$\astep{\eval{\rho_{n+1}}(\encap{H}(p_{n+1}))}{\gact{\rho}
  {\alpha'}}{\eval{\rho'}(p')}$}
for some $p'$ and $\rho'$. 
This means that it cannot even be established whether the enabledness 
condition of $p$ is too strong for $\encap{H}(p)$.

In the rest of this section, we describe the behaviour of a very simple 
system by a closed \deACPei\ term and reason about how this system 
changes data with the adapted axioms and rules of \RG.
The assumptions about $\gD$ are as in Section~\ref{sect-example}.
Moreover, it is assumed that $i$ is a flexible variable from $\FlexVar$.
The behaviour of the very simple system concerned is described by the 
closed \deACPei\ \linebreak[2] term 
$(i > 0 \gc \ass{i}{2}) \parc \ass{i}{1}$. 

In \RG, the asserted process with enablednes-condition
\begin{ldispl}
\langle 
 i = \prevvar{i},\True, 
 i = \prevvar{i} \Lor (\prevvar{i} = 0 \Land i = 1) \Lor
                      (\prevvar{i} > 0 \Land i = 2)\, \rangle : 
\qquad \\ \hfill
\hlassproc{i = 0}{(i > 0 \gc \ass{i}{2}) \parc \ass{i}{1}}{i = 2}
\end{ldispl}%
is true in the sense of weakly total correctness.
Below, it is sketched how this asserted process can be derived by means 
of the adapted axioms and rules of \RG.

We derive, using the assignment action rule and the guarded command 
rule, the following asserted process:
\begin{ldispl}
\langle i = \prevvar{i} \Lor (\prevvar{i} = 0 \Land i = 1),i > 0,
        i = \prevvar{i} \Lor (\prevvar{i} > 0 \Land i = 2)\, \rangle :
\qquad \\ \hfill
\hlassproc{i = 0}{i > 0 \gc \ass{i}{2}}{i = 2}
\end{ldispl}%
and we derive, using the assignment action rule and the consequence 
rule, the following asserted process:
\begin{ldispl}
\langle i = \prevvar{i} \Lor (\prevvar{i} > 0 \Land i = 2),\True,
        i = \prevvar{i} \Lor (\prevvar{i} = 0 \Land i = 1)\, \rangle :
\phantom{\,.} \qquad \\ \hfill
\hlassproc{i = 0}{\ass{i}{1}}{i = 1 \Lor i = 2}\;.
\end{ldispl}%
From these two asserted processes, we derive, using the parallel 
composition rule, the following asserted process:
\begin{ldispl}
\langle 
 i = \prevvar{i},\True, 
 i = \prevvar{i} \Lor (\prevvar{i} = 0 \Land i = 1) \Lor
                      (\prevvar{i} > 0 \Land i = 2)\, \rangle : 
\phantom{\,.} \qquad \\ \hfill
\hlassproc{i = 0}{(i > 0 \gc \ass{i}{2}) \parc \ass{i}{1}}{i = 2}\;.
\end{ldispl}%
Because the enabledness-condition in the above asserted process is 
$\True$, the process $(i > 0 \gc \ass{i}{2}) \parc \ass{i}{1}$ is 
deadlock-free if the value of $i$ equals $0$ at its start and each step 
taken by processes taken place in parallel with it does not change the 
value of $i$.

\section{Concluding Remarks}
\label{sect-conclusions}

There is extensive literature on rely/guarantee logics in which the 
judgments concern programs.
In \RG, the rely/guarantee logic developed in the current paper, the 
judgments instead concern processes that can be represented by a term 
from an imperative process algebra.
In order to build on earlier work on rely/guarantee logics, the 
operational semantics of process terms underlying \RG\ is of the kind 
commonly used for programs.
This operational semantics has been extracted from the operational 
semantics underlying the axiom system of \deACPei.
It should be mentioned that the work presented in this paper mainly 
builds on the earlier work on rely/guarantee logic presented 
in~\cite{XRH97a}.

The use of the auxiliary variable rule of \RG\ actually breaks 
compositionality.
Therefore, it is discouraged to use auxiliary variables.
There are two problems that lead to the use of auxiliary variables 
anyway: an expressiveness problem and a completeness problem.
For either problem, it is an open question whether there exists a 
compositional alternative in the setting of \RG\ that solves the 
problem. 
The former problem has been studied in the setting of a rely/guarantee 
logic in which the judgments concern programs rather than processes and
the latter problem has been studied in the setting of a Hoare logic
rather than rely/guarantee logics.
I consider studies of these problems in the setting of \RG\ --- a 
rely/guarantee logic in which the judgments concern processes --- 
interesting options for future work.

The adapted encapsulation rule of \RG\ for asserted processes with 
enabledness-condition has a side condition which makes the rule 
inapplicable in most practical cases while encapsulation is necessary in
virtually all application of \deACPei.
It is an open question whether this problem can be solved by extending 
\deACPei\ with ``encapsulating parallel composition'' operators 
$\parc_H$.

\appendix

\section{Proof Outlines}
\label{appendix-proofs}
In this appendix, the routine proofs of 
Proposition~\ref{proposition-equiv-deACPei},
Proposition~\ref{proposition-congr-deACPei}, and
Theorem~\ref{theorem-soundness-deACPei} are outlined.

\subsubsection*{Proof of Proposition~\ref{proposition-equiv-deACPei}}
It has to be shown that $\bisim$ is reflexive, symmetric, and 
transitive.

The identity relation $I$ is a bisimulation witnessing $t \bisim t$ for 
all $t \in \ProcTerm$.
Hence, $\bisim$ is reflexive.

Let $t_1, t_2 \in \ProcTerm$ be such that $t_1 \bisim t_2$, and
let $R$ be a bisimulation witnessing $t_1 \bisim t_2$.
Then $R^{-1}$ is a bisimulation witnessing $t_2 \bisim t_1$.
Hence, $\bisim$ is symmetric.

Let $t_1, t_2, t_3 \in \ProcTerm$ be such that $t_1 \bisim t_2$ and
$t_2 \bisim t_3$, 
let $R$ be a bisimulation witnessing $t_1 \bisim t_2$, and 
let $S$ be a bisimulation witnessing $t_2 \bisim t_3$.
Then $R \circ S$ is a bisimulation witnessing 
$t_1 \bisim t_3$.%
\footnote{We write $R \circ S$ for the composition of $R$ with $S$.}
Hence, $\bisim$ is transitive.
\qed

\subsubsection*{Proof of Proposition~\ref{proposition-congr-deACPei}}
In~\cite{BV93a}, an SOS rule format, called the path format, is 
introduced.
Theorem~5.4 from that paper expresses that if a set of transition 
rules make up a well-founded transition system specification in path 
format then the standard notion of bisimulation equivalence is a 
congruence with respect to all operators involved.
It is easy to establish that the set of transition rules for \deACPei\ 
make up a well-founded transition system specification in path format.
Because $\bisim$ is an instance of the standard notion of bisimulation 
equivalence, the proposition follows immediately.
\qed

\subsubsection*{Proof of Theorem~\ref{theorem-soundness-deACPei}}
Because ${\bisim}$ is a congruence with respect to all operators from 
the signature of \deACPei, it is sufficient to prove the theorem for 
all closed substitution instances of each axiom of \deACPei.
We will loosely say that a relation contains all closed substitution 
instances of an equation if it contains all pairs $(t,t')$ such that 
$t = t'$ is a closed substitution instance of the equation.

For each equational axiom of \deACPei, a bisimulation $R$ witnessing 
$t \bisim t'$ for all closed substitution instances $t = t'$ of the 
axiom can be constructed.
It is easy to check that the following constructions do indeed produce 
witnesses:
\begin{itemize}
\item
if the axiom is one of the axioms A7, CM2E, CM5E, CM6E, GC2 or an 
instance of one of the axiom schemas D0, D2, GC3, V0, CM7Db--CM7Df, 
then $R$ is the relation that consists of all closed substitution 
instances of the axiom concerned;
\item
if the axiom is one of the axioms A1--A6, A8, A9, CM4, CM8--CM9, 
BKS1, BKS5, GC1 or an instance of one of the axiom schemas CM3, CM7, D1, 
D3, D4, GC4--GC11, V1--V5, CM7Da, then $R$ is the relation that consists 
of all closed substitution instances of the axiom concerned and the 
equation $x = x$;
\item
if the axiom is CM1E, then $R$ is the relation that consists of all 
closed substitution instances of CM1E, the equation 
$x \parc y = y \parc x$, and the equation $x = x$;
\end{itemize}
For the only conditional equational axiom of \deACPei, viz.\ RSP*E, a 
bisimulation $R$ witnessing $t \bisim t' \iter t''$ for all closed 
substitution instances $t = t' \iter t''$ of the consequent of RSP*E for 
which $\encap{\Act}(t) \bisim \dead$ and 
$t \bisim t' \seqc t \altc t''$ can be constructed.
It is easy to check that the following construction does indeed produce 
a witness: $R$ is the relation that consists of all closed substitution 
instances $t = t' \iter t''$ of the consequent of RSP*E for which 
$\encap{\Act}(t) \bisim \dead$ and $t \bisim t' \seqc t \altc t''$ 
and all closed substitution instances of the equation $x = x$. 
\qed

\bibliographystyle{splncs04}
\bibliography{PA}

\end{document}